\pdfoutput=1
\documentclass[aps,prb,reprint,showpacs,twocolumn, superscriptaddress,preprintnumbers, amsmath, epsfig,floatfix]{revtex4-1}
\usepackage{graphicx}
\usepackage{amsmath}
\usepackage{amssymb}
\usepackage{amsfonts}
\usepackage{dcolumn}
\usepackage{dsfont}
\usepackage{latexsym}
\usepackage{rotating}
\usepackage{color}
\usepackage{latexsym}
\usepackage{bbm}
\usepackage{subfigure}
\usepackage{float}
\usepackage{epsfig}
\usepackage{psfrag}
\usepackage{natbib}
\usepackage{bm}
\usepackage{amsthm}
\usepackage{eucal}
\usepackage{mathrsfs}
\usepackage{url}
\usepackage{braket}
\usepackage{mathtools}
\usepackage{amsmath,amssymb}
\usepackage{epsfig}
\usepackage{amsmath,amssymb}
\usepackage{hyperref}
\usepackage{soul}

\usepackage{color}

\usepackage{hyperref}
\hypersetup{
	colorlinks=true,final=true,
	linkcolor=blue,
	citecolor=blue,
	filecolor=blue,
	urlcolor=blue,
}

\begin{document}

\title{Engineering room-temperature multiferroicity in Bi and Fe codoped BaTiO$_3$}

\author{Pratap Pal}
\affiliation{Department of Physics, Indian Institute of Technology Kharagpur, West Bengal 721302, India}
\author{Tapas Paramanik}
\affiliation{Department of Physics, Indian Institute of Technology Kharagpur, West Bengal 721302, India}
\affiliation{Department of Physics, School of Sciences, National Institute of Technology Andhra Pradesh, Tadepalligudem 534101, India}
\author{Krishna Rudrapal}
\affiliation{Advanced Technology Development Centre, Indian Institute of Technology Kharagpur, West Bengal 721302, India}
\author{Supriyo Majumder}
\affiliation{UGC-DAE Consortium for Scientific Research, University Campus, Khandwa Road, Indore 452001, India}
\author{Satish Yadav}
\affiliation{UGC-DAE Consortium for Scientific Research, University Campus, Khandwa Road, Indore 452001, India}
\author{Sudipta Mahana}
\affiliation{Rajdhani College, Baramunda square, Bhubaneswar 751003, India}
\author{Dinesh Topwal}
\affiliation{Institute of Physics, Sachivalaya Marg, Bhubaneswar 751005, India}
\affiliation{Homi Bhabha National Institute, Training School Complex, Anushakti Nagar, Mumbai 400085, India}
\author{Ram Janay Choudhary}
\affiliation{UGC-DAE Consortium for Scientific Research, University Campus, Khandwa Road, Indore 452001, India}
\author{Kiran Singh}
\affiliation{UGC-DAE Consortium for Scientific Research, University Campus, Khandwa Road, Indore 452001, India}
\affiliation{Dr. B. R. Ambedkar National Institute of Technology, Jalandhar 144011, India}
\author{Ayan Roy Chaudhuri}
\affiliation{Advanced Technology Development Centre, Indian Institute of Technology Kharagpur, West Bengal 721302, India}
\affiliation{Materials Science Centre, Indian Institute of Technology Kharagpur, West Bengal 721302, India}
\author{Debraj Choudhury}
\email{debraj@phy.iitkgp.ac.in}
\affiliation{Department of Physics, Indian Institute of Technology Kharagpur, West Bengal 721302, India}

\begin{abstract}
	
Fe doping into BaTiO$_3$, stabilizes the paraelectric hexagonal phase in place of the ferroelectric tetragonal one [P. Pal {\it{et}} {\it{al.}} {\it{Phys. Rev. B}}, {\bf{101}}, 064409 (2020) \cite{PPal2020}]. We show  that simultaneous doping of Bi along with Fe into BaTiO$_3$ effectively enhances the magnetoelectric (ME) multiferroic response (both ferromagnetism and ferroelectricity) at room-temperature, through careful tuning of Fe valency along with the controlled-recovery of ferroelectric-tetragonal phase. We also report systematic increase in large dielectric constant values as well as reduction in loss tangent values with relatively moderate temperature variation of dielectric constant around room-temperature with increasing Bi doping content in Ba$_{1-x}$Bi$_{x}$Ti$_{0.90}$Fe$_{0.10}$O$_{3}$ (0$\leq${\it{x}}$\leq$0.10), which makes the higher Bi-Fe codoped sample ({\it{x=0.08}}) promising for the use as room-temperature high-$\kappa$ dielectric material. Interestingly, {\it{x=0.08}} (Bi-Fe codoped) sample is not only found to be ferroelectrically ($\sim$20 times) and ferromagnetically ($\sim$6 times) stronger than {\it{x=0}} (only Fe-doped) at room temperature, but also observed to be better insulating (larger bandgap) with indirect signatures of larger ME coupling as indicated from anomalous reduction of magnetic coercive field with decreasing temperature. Thus, room-temperature ME multiferroicity has been engineered in Bi and Fe codoped BTO (BaTiO$_3$) compounds.

\end{abstract}

\maketitle

\section{Introduction}

 Room-temperature ME multiferroic materials are extremely promising for future technological applications \cite{NAHill2000,DKhomskii2009}. In this regard, efforts toward turning BaTiO$_3$ (a strong proper ferroelectric \cite{RECohen1992}) into multiferroic material by Fe \cite{BXu2009,PPal2020} (or other transition metals like Mn \cite{YHLin2009}, Co \cite{LBLuo2009} etc.) doping has attracted a lot of attention. However, the observed multiferroicity in Fe-doped BTO is seen to be of mixed phase kind where the ferromagnetism arises from the majority paraelectric-hexagonal phase and weak ferroelectricity due to a minority tetragonal phase  \cite{PPal2020}. Thus, in such a case the ME coupling is naturally supposed to be weak due to the independent origin of two ferroic orders at two different structural units. So, the necessity is to stabilize the tetragonal phase, even in this magnetically Fe-doped BTO compound, such that both ferroelectricity and ferromagnetism can emerge from the same tetragonal phase, which may lead to better ME coupling along with pronounced ferroelectricity. It is important to note that for a material to be a potential room-temperature multiferroic, it is desirable to have (a) a strong ferroelectric polarization, (b) a high ferromagnetic remanent moment, (c) a low leakage-current density (i.e. highly insulating) along with (d) ME coupling between the two ferroic-orders \cite{NASpaldin2019}.

 %_________fig1

 \begin{figure}[h!]
 	\begin{center}
 		\vspace*{-0.15 in}
 		\hspace*{-0.0in} \scalebox{0.3}{\includegraphics{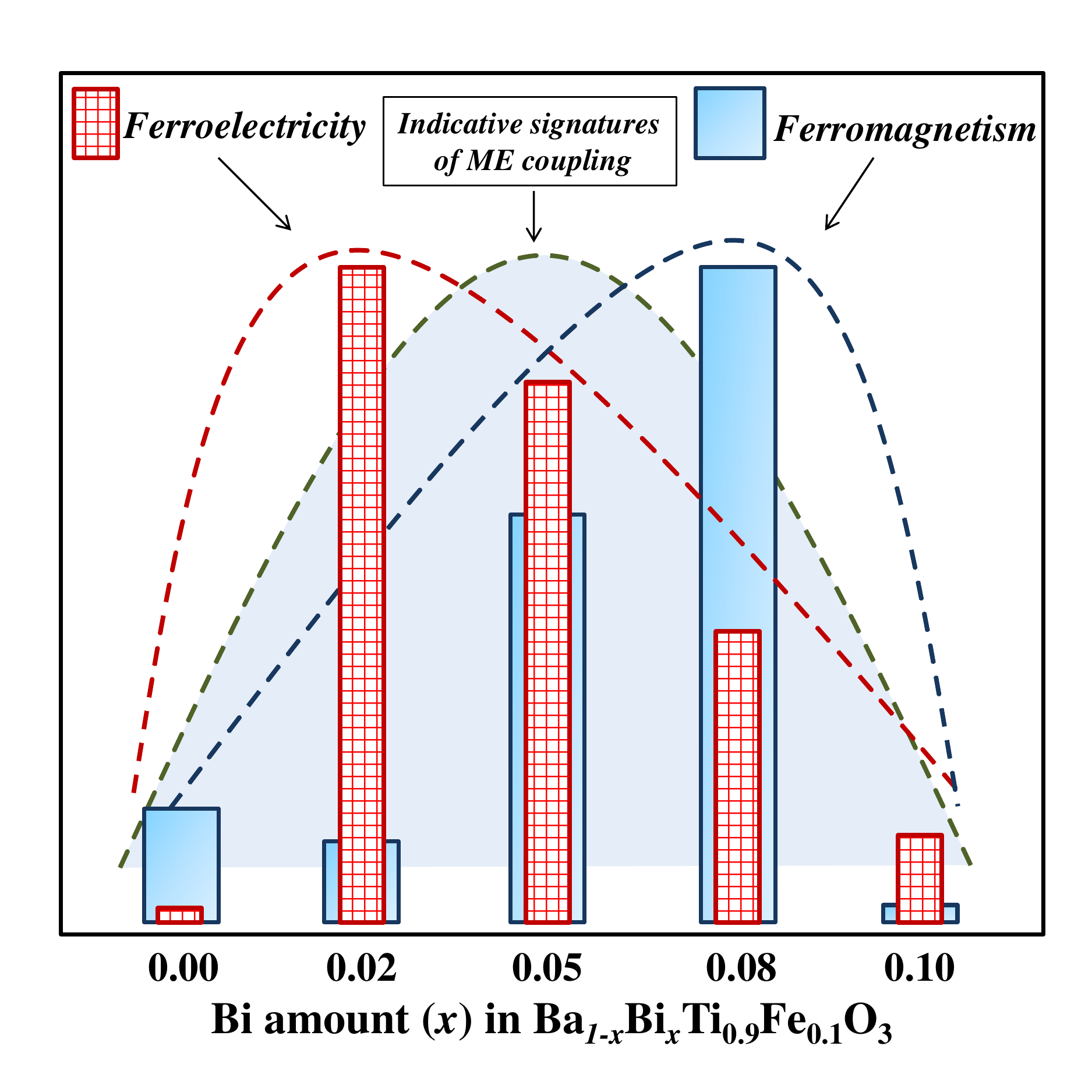}}
 		\vspace*{-0.15 in}\caption{(Color online) Schematic representation of the multiferroic response from Ba$_{1-x}$Bi$_{x}$Ti$_{0.90}$Fe$_{0.10}$O$_{3}$ compounds. The best ferroelectric and magnetic responses are obtained around {\it{x=0.02}} and {\it{x=0.08}} regions respectively, while the indicative signature of ME coupling becomes the strongest for the intermediate {\textit{x=0.05}} composition (bar height represents the relative strength of the observed physical property).}\label{Phasediagram}
 	\end{center}
 \end{figure}

With these goals, we examine polycrystalline Ba$_{1-x}$Bi$_{x}$Ti$_{0.90}$Fe$_{0.10}$O$_{3}$ compounds (0$\leq${\it{x}}$\leq$0.10) and show that with increasing Bi doping content not only the paraelectric-hexagonal phase of {\it{x=0}} (only Fe-doped BTO) gets destabilized (with concomitant recovery of ferroelectric-tetragonal phase), but also there is a surprising and significant enhancement of room-temperature ferromagnetism. Interestingly, we find that the observed ferroelectricity is strongest around {\it{x=0.02}} and ferromagnetism around {\it{x=0.08}}, while the intermediate {\it{x=0.05}} shows indirect signature of most robust ME-coupling (see Fig.\ref{Phasediagram}).

\section{Experimental details}

All phase pure (related to BTO hexagonal and tetragonal phases only) polycrystalline samples which were prepared via solid state reactions (for details refer to ref.\cite{PPal2020}) were characterized using XRD and Raman measurements. Dielectric and magnetic measurements were performed using a Keysight E4980A LCR meter and SQUID-VSM respectively. Electron Paramagnetic Resonance (EPR) measurements were carried out using Bruker spectrometer. Intrinsic ferroelectric properties have been investigated using ferroelectric PUND (Positive Up Negative Down) and temperature dependent pyroelectric measurements \cite{PPal2020}. XANES and XPS measurements were carried out at P-65 beamline in PETRA-III synchrotron source, DESY, Hamburg, Germany and using an in-house PHI 5000 Versaprobe-II spectrometer respectively, while the diffused reflectance spectra were acquired using a Cary 5000 UV-Visible near-infrared spectrophotometer.

\section{Results and discussions}

\subsection{Tuning of room-temperature ferroelectricity}

\begin{figure}[t]
	
	\vspace*{-0.1 in}
	\hspace*{0.15in} \scalebox{0.37}{\includegraphics{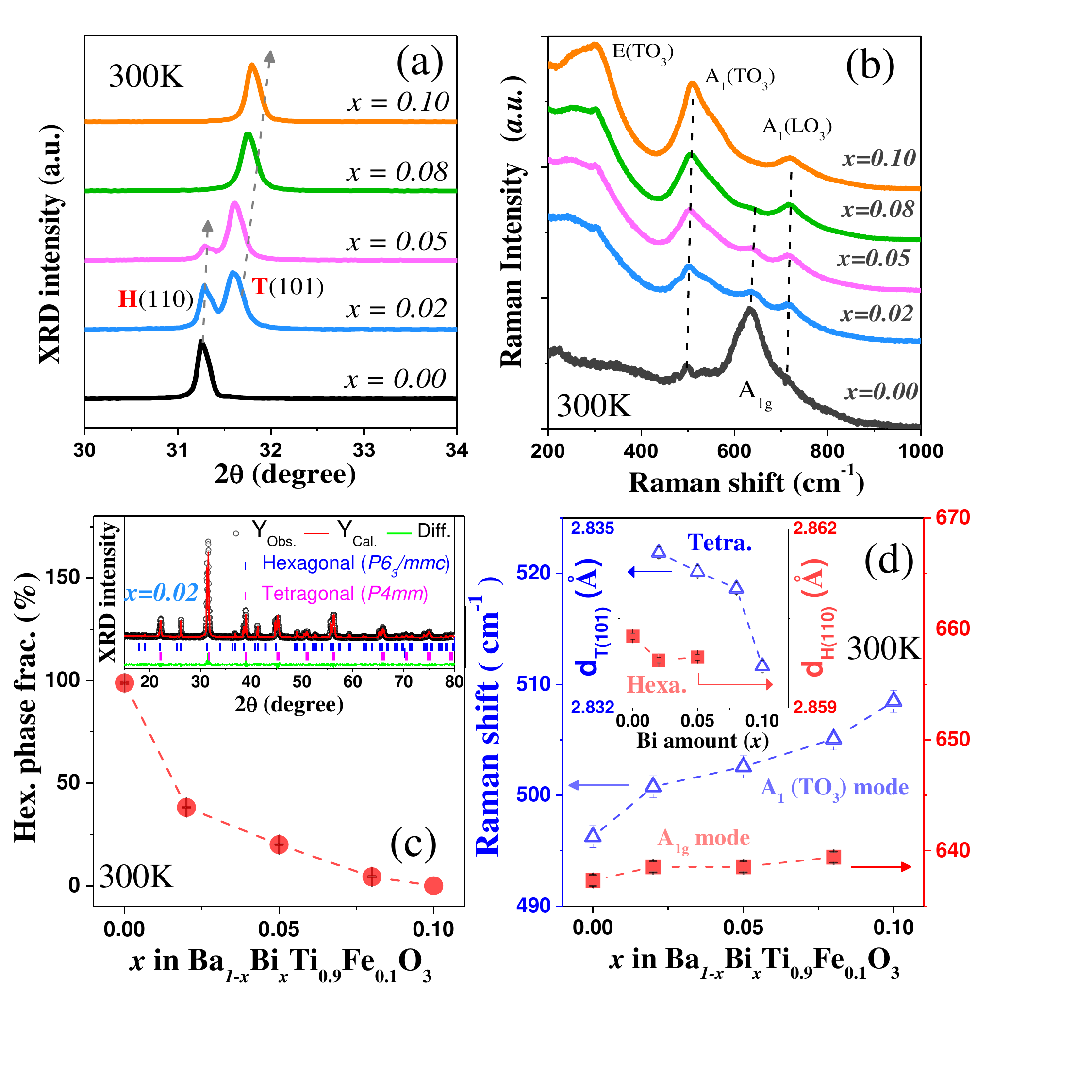}}
	\vspace*{-0.5 in}\caption{(Color online) (a) Room temperature XRD spectra of Ba$_{1-x}$Bi$_{x}$Ti$_{0.90}$Fe$_{0.10}$O$_{3}$, (b) corresponding room-temperature Raman spectra, (c) variation of the hexagonal phase fraction (\%). Inset to (c) shows Rietveld refinement for {\it{x}}=0.02 and (d) Raman shifting of tetragonal A$_1$(TO$_3$) mode and hexagonal A$_{1g}$ mode respectively, while its inset shows variation of interplannar spacings of hexagonal H(110) and tetragonal T(101) planes respectively. }\label{TuningHexagonality}
	
\end{figure}

\begin{figure}[t]
	\vspace*{-00.2 in}
	\hspace*{0.15in}\scalebox{0.37}{\includegraphics{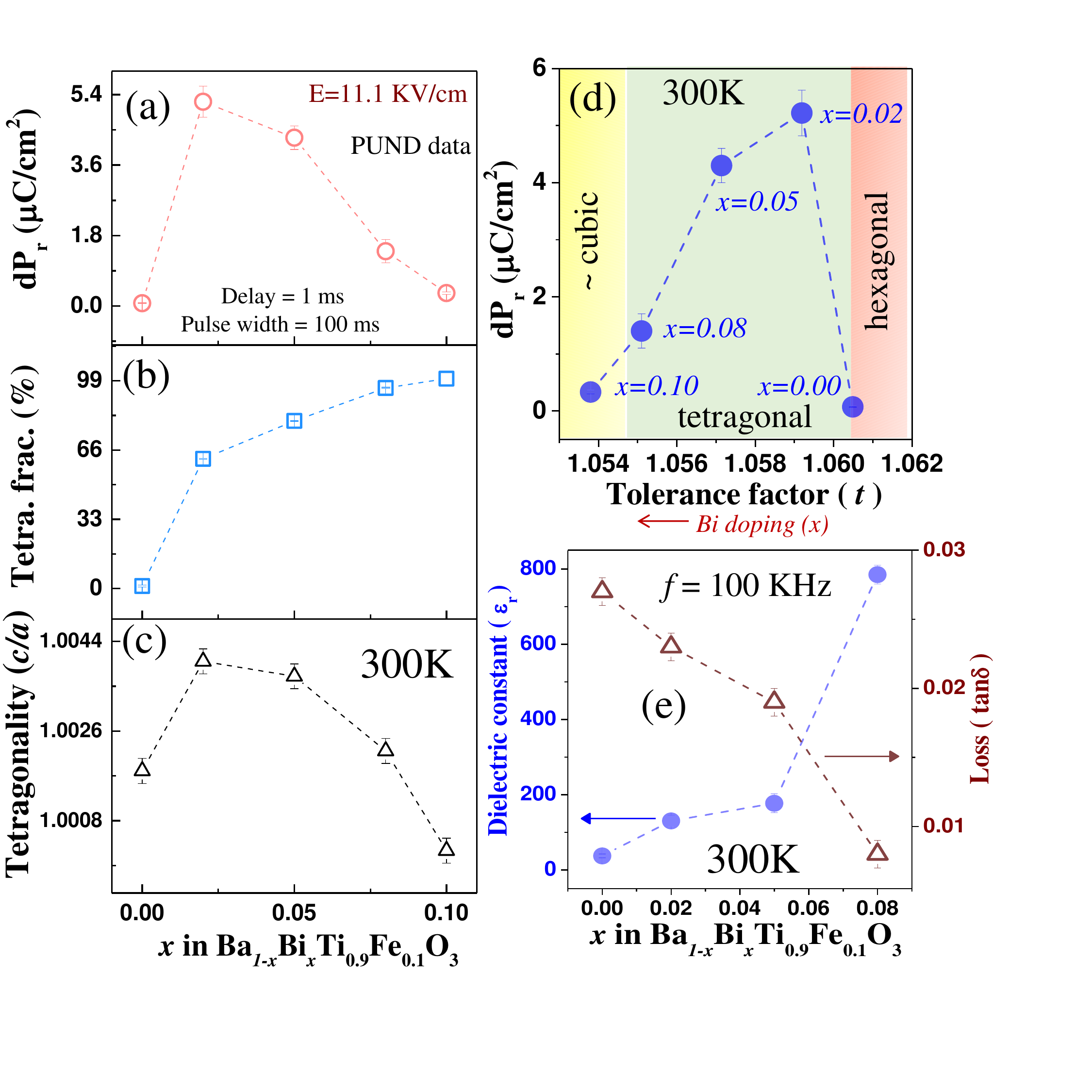}}\vspace*{-0.4in}\caption{(Color online) (a) variation of remanent-ferroelectric polarization at room temperature, obtained from PUND measurements, (b) and (c) show the variation of tetragonal phase fraction(\%) and sample tetragonality ({\it{c/a}}) respectively. (d) Variation of remanent polarization with the sample tolerance factor, and (e) variation of dielectric constant and loss tangent values at room temperature with Bi doping amount.}\label{Ferroelectricity}
	
\end{figure}

Fig.\ref{TuningHexagonality}(a) covers the 2$\theta$ range in XRD spectrum which includes the strongest diffraction peaks from the hexagonal [H(110)] and the tetragonal [T(101)] BTO phases [also, note Fig. \ref{BBTFO-XRD} of the supplementary material section below]. While {\it{x=0}} crystallizes in the hexagonal phase, with increasing {\it{x}}, the tetragonal phase gets fully recovered as observed from the systematic increase in tetragonal T(101) peak intensity and concomitant decrease in hexagonal H(110) peak intensity \cite{DHKim2016}. The XRD observations of tetragonal phase recovery are further validated from Raman spectroscopy as shown in Fig.\ref{TuningHexagonality}(b), where clear evolution of tetragonal Raman modes from the majority hexagonal phase with increasing Bi concentration has been observed \cite{CHPerry1965,HaMNguyen2011}. The amount of hexagonal [shown in Fig.\ref{TuningHexagonality}(c)] and tetragonal phase [shown in Fig.\ref{Ferroelectricity}(b)] fractions present in each of these samples are determined from Rietveld refinement of their room-temperature XRD spectra incorporating both hexagonal ({\it{P6$_3$/mmc}}) and tetragonal ({\it{P4mm}}) phases [an example is shown in the inset of Fig.\ref{TuningHexagonality}(c) for {\it{x=0.02}}, for further details note Fig. \ref{Refinement-details} of supplementary section below]. It is interesting to note that there is strong shifting of the tetragonal T(101) XRD peak to higher angles with Bi doping, however, such is not the case for the hexagonal H(110) as shown in Fig.\ref{TuningHexagonality}(a). Similarly, no significant change in d$_{H(110)}$ [spacing between H(110) planes] is observed beyond {\it{x=0.02}}, whereas d$_{T(101)}$ systematically gets reduced with increasing Bi doping concentration, as shown in the inset to Fig.\ref{TuningHexagonality}(d). The reduction of d$_{T(101)}$ is driven by the substitution of Ba$^{2+}$ ions by relatively smaller Bi$^{3+}$ ions. Further, we note the monotonic shifting of the tetragonal Raman A$_1$(TO$_3$) mode towards higher energy with Bi doping, which is, however, not quite prominent in case of the hexagonal Raman A$_{1g}$ mode as shown in Fig.\ref{TuningHexagonality}(d). Thus, both XRD and Raman analyses suggest a maximum doping percentage of Bi at the Ba site of the hexagonal BTO.

%_________fig2

 %_________fig3

Next, to study the intrinsic ferroelectric properties, we have employed the PUND \cite{BRajeswaran2012,ARChaudhuri2010} technique (for details see ref.\cite{PPal2020}). In Fig.\ref{Ferroelectricity}(a) we see that from {\it{x=0}} there is a sudden increase in the ferroelectric polarization value for {\it{x=0.02}} and then it continues to decrease up to {\it{x=0.10}} [also, refer to Fig. \ref{PUND} of the supplementary material below]. However, such nonmonotonic trend of polarization-change can not be understood from the monotonic evolution of tetragonal phase fraction (\%) as seen in Fig.\ref{Ferroelectricity}(b). To investigate such a behavior, we have calculated the sample tetragonality ({\it{c/a}}), which is shown in Fig.\ref{Ferroelectricity}(c). Here, we note that the tetragonality, which is found to be maximum for {\it{x=0.02}}, exhibits the same trend as observed for polarization change in Fig.\ref{Ferroelectricity}(a). Therefore, the tetragonality, measure of the amount of Ti-off centric distortion in the corresponding TiO$_6$ cage \cite{RECohen1992} (see Fig. \ref{Octahedral-distortion} of the supplementary material below), overrides the amount of tetragonal phase percentage present in regards to the observed ferroelectricity in these systems \cite{PPal2020}. Such observations were further validated from temperature dependent pyrocurrent measurements (see Fig. \ref{Pyro} and note the discussions in subsection-\textbf{C} of the supplementary material below).

\begin{table}[t]
	\centering
	\caption{ Dielectric constant ( $\epsilon$$_r$ ), dielectric loss (D), and the temperature coefficient of dielectric constant TC$_\epsilon$ (which is averaged from 275 K to 325 K temperature window) measured at a frequency of  $\sim$100 kHz. Here, only values of SiO$_2$ and SrTiO$_3$ are taken from previous reports, reproduced from [D. Choudhury \textit{et al.} Appl. Phys. Lett. \textbf{96}, 162903 (2010) \cite{DChoudhury2010}], with the permission of AIP Publishing.}\label{Dielectric}
	\vspace*{0.5cm}
	\begin{tabular}{|c|c|c|c|}
		\hline
		
		Sample & Dielectric  & Loss (D) & Avg. TC$_\epsilon$  \\
		&constant ($\epsilon$$_r$) &  & (ppm/K) \\
		& 300K & 300 K & at (300$\pm$25)K \\
		\hline

		SiO$_2$ & 3.7 & 0.0015 & ... \\

		SrTiO$_3$  & 285 & 0.006 & -2400  \\
	
	    BaTiO$_3$ & 1626 & 0.015 & -799 $\mp$ 408\%  \\
	
		{\it{x=0}} & 37 & 0.027 & -541 $\mp$ 8\% \\
	
		{\it{x=0.05}} & 177 & 0.019 & 2358 $\pm$ 17\%   \\
	
		{\it{x=0.08}} & 785 & 0.008 & 2262 $\mp$ 14\%   \\

		\hline
	\end{tabular}
	
\end{table}

 Now, to have a holistic view, we have plotted the remanent polarization value (dPr) versus sample tolerance factor (which can strongly influence structural instabilities of such perovskite oxides \cite{ZLi2016}) as shown in Fig.\ref{Ferroelectricity}(d), using Fe and Ti valencies determined from the Fe-K edge XANES and XPS spectroscopic studies \cite{PPal2020} (for details, refer to subsections \textbf{E}-\textbf{G} and note Fig. \ref{XANES} and Fig. \ref{XPS} of the supplementary material below). Interestingly, there is a window of tolerance factor around which ferroelectric response is optimum in these systems, useful for achieving optimized multiferroic response.

\subsection{Bi-Fe codoped BTO: promising as high-$\kappa$ dielectric}

 Interestingly, we find systematic increase in large dielectric constant values as well as reduction in loss tangent values with increasing Bi doping content [see Fig.\ref{Ferroelectricity}(e)], which makes the higher Bi-Fe codoped system ({\it{x=0.08}}) promising for the use as room-temperature high-$\kappa$ dielectric material since, it exhibits a unique combination of often contradicted physical properties i.e. very large dielectric constant, extremely low dielectric loss and moderate temperature dependence of dielectric constant values around room-temperature [TC$_\epsilon$ = (1/$\epsilon$$_r$)$\times$ ({$\partial${$\epsilon$$_r$}}/{$\partial${T}})] (see table-\ref{Dielectric}) \cite{DChoudhury2010,AIKingon2000}.  While undoped BaTiO$_3$ also exhibits a very high dielectric constant and low loss values, but the corresponding $\mid$TC$_\epsilon$$\mid$ value changes a lot within 50 K temperature window around room-temperature rendering it unsuitable as a high-$\kappa$ dielectric material. Interestingly, the dielectric response of {\it{x=0.08}} (i.e. higher dielectric constant value, comparable loss and
 $\mid$TC$_\epsilon$$\mid$ values) seems better compared to SrTiO$_3$, very well-known as a high-$\kappa$ dielectric material \cite{KEisenbeiser2000}.

\subsection{Tuning of room-temperature ferromagnetic properties}

\begin{figure}[t]
	\vspace*{-0.1 in}
	\hspace*{0.1 in} \scalebox{0.37}{\includegraphics{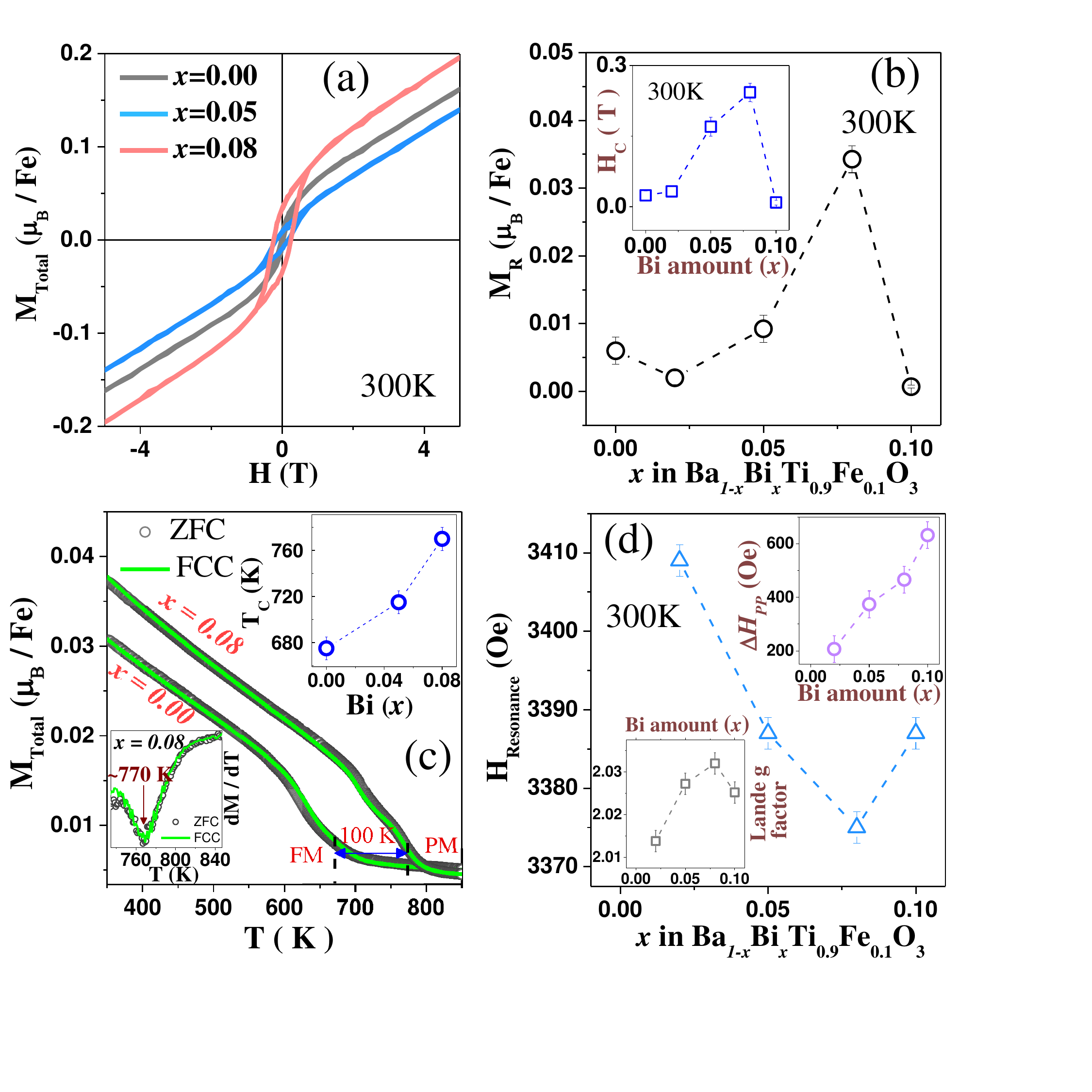}}
	\vspace*{-0.6 in}\caption{ (Color online) (a) Room temperature MH plots. (b) variation of remanent magnetization (M$_R$). Inset to it shows the corresponding change in coercive field (H$_C$). (c) Temperature dependent magnetization behavior for applied field of 0.5 T, where bottom and top insets respectively shows dM/dT vs. T plot and the change in ferromagnetic T$_C$ with Bi doping concentration. (d) and its top inset show variation of resonance magnetic field and H$_{PP}$ (peak to peak line-width) as determined from EPR spectra, while its bottom inset shows corresponding change in Lande g-factor.}\label{Ferromagnetic-EPR}
\end{figure}

To investigate the magnetic properties, first we discuss the room-temperature MH data as shown in Fig.\ref{Ferromagnetic-EPR}(a), which exhibit finite ferromagnetic loops, however do not saturate, indicating presence of some paramagnetic contributions \cite{DKarmakar2007}. Thus, we consider the intrinsic ferromagnetic remanent moment (M$_{R}$) which is seen to monotonically increase from {\it{x=0}} up to {\it{x=0.08}} (having $\sim$6 times enhanced magnetization) and then decreases for {\it{x=0.10}} as shown in Fig.\ref{Ferromagnetic-EPR}(b). Such a non-monotonic trend is also reflected even in case of coercive field [see inset to Fig.\ref{Ferromagnetic-EPR}(b)]. Temperature dependent magnetization measurements display clear ferromagnetic to paramagnetic phase transition and the Curie temperature (T$_C$) [as estimated from dM/dT vs. T plot, shown in the bottom inset to Fig.\ref{Ferromagnetic-EPR}(c)] is found to increase systematically with Bi doping concentration as shown in Fig.\ref{Ferromagnetic-EPR}(c) and its top inset. To further investigate the above trend of magnetization in the microscopic scale, we employ EPR spectroscopy, which is shown in Fig. \ref{EPR} (also, note subsection-\textbf{J}) of the supplementary material section, where we see marked changes in line-shape with Bi doping. First, we have determined peak-to-peak line width ($\Delta$H$_{PP}$) [see top inset of Fig.\ref{Ferromagnetic-EPR}(d)], which increases monotonically with Bi doping, indicating increasing Fe-Fe exchange interaction \cite{SDBhame2005,TChakraborty2011}. Subsequently, we see that the resonance magnetic field monotonically decreases up to {\it{x=0.08}} and then slightly increases for {\it{x=0.10}}. Such observations strongly indicate that the strength of ferromagnetic interaction gets enhanced up to {\it{x=0.08}}, followed by a reduction for {\it{x=0.10}}, possibly due to some antiferromagnetic interaction \cite{JGutierrez2000,SDBhame2005}. Even, the Lande g-factor for all these compounds, as shown in the bottom inset of Fig.\ref{Ferromagnetic-EPR}(d), shows similar trend like the remanent magnetization value. Such enhancement of ferromagnetism, however, can not be understood from the frame work of increasing oxygen vacancy content \cite{TChakraborty2011} or increasing hexagonal phase fractions, as both are simultaneously and monotonically reduced with increasing Bi doping concentration \cite{PPal2020}. Such trends of magnetic response cannot be understood from the possible presence of any extrinsic contributions, like from BiFeO$_3$, Fe$_2$O$_3$ or Fe$_3$O$_4$ related impurity phases, which, however, could not be detected through detailed XRD analyses, as discussed in the subsection-\textbf{M} of the supplementary material (also, note references \cite{JLu2010,ASTeja2009,YIshikawa1957,SAkimoto1957}). These results suggest an alternative source of ferromagnetism for higher {\it{x}} members beyond the hexagonal BTO phase, that can likely be the tetragonal BTO phase (Fe$^{4+}$-O-Fe$^{3+}$, FM superexchange interaction), as beyond some doping limit both Bi and Fe tend to go into the tetragonal phase (for schematic visualization of superexchange, see sub-section \textbf{I} and note Fig. \ref{Tetragonal-superexchange} and Fig. \ref{Hexagonal-superexchange} of the supplementary material below).

%_________fig4

%_________fig5
\begin{figure}[t]
	
	\vspace*{-0.1 in}
	\hspace*{0.08 in} \scalebox{0.38}{\includegraphics{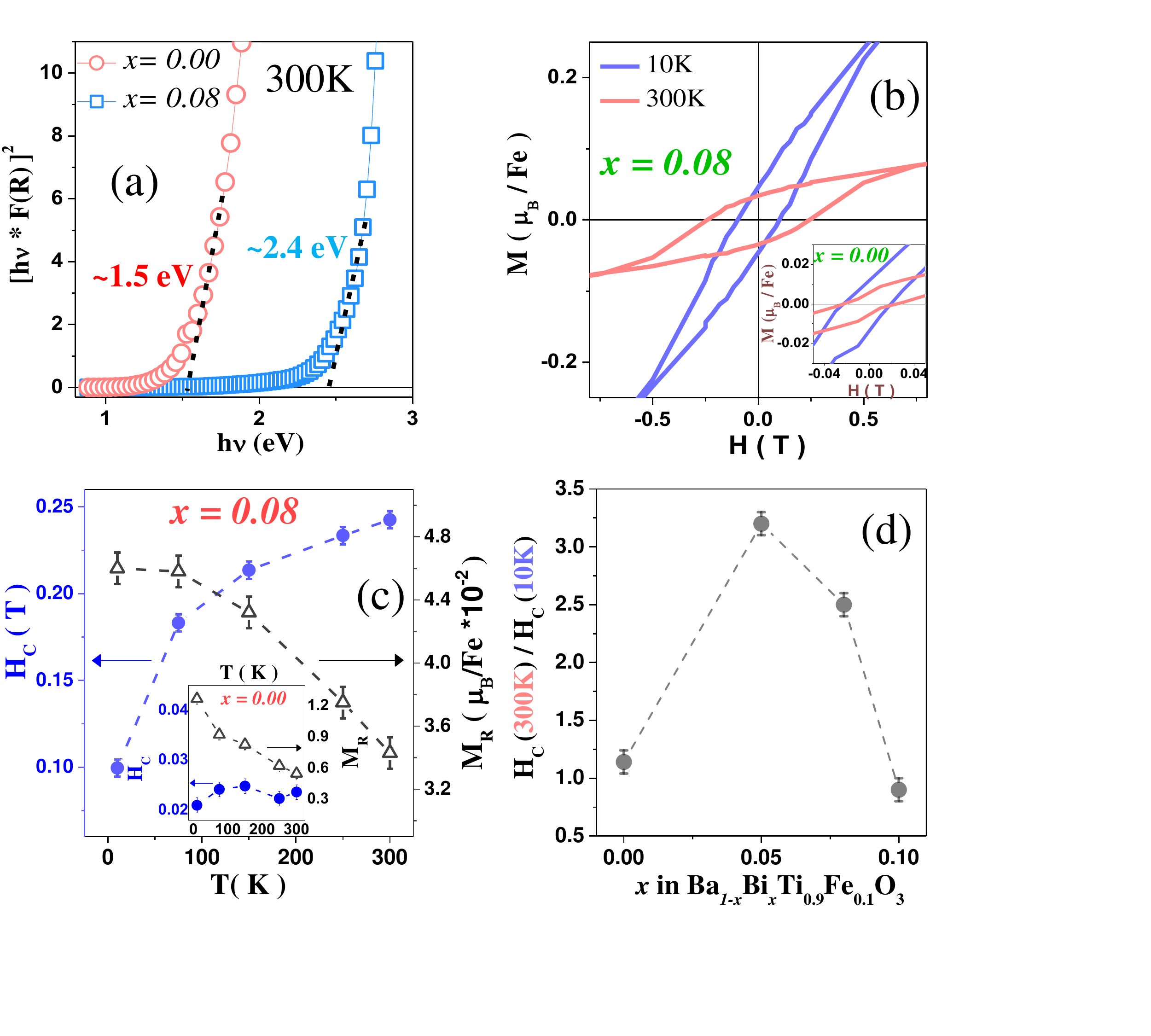}}
	\vspace*{-0.6 in}\caption{ (Color online) (a) Kubelka-Munk plots calculated from diffused reflectance spectroscopy, (b) MH plots of {\it{x=0.08}} at 300 K and 10 K, where its inset shows the corresponding MH plots of {\it{x=0}}. (c) and its inset show variations of coercive field (H$_C$) and remanent magnetization (M$_R$) with temperature for {\it{x=0.08}} and {\it{x=0}} respectively. (d) variation of the ratio of coercive field (H$_C$) at 300 K and 10 K with Bi doping.}\label{MultiferroicityImproved}
	
\end{figure}

\subsection{Enhanced room-temperature multiferroicity in Bi-Fe codoped BTO}

 The {\it{x=0.08}} compound, in addition to possessing $\sim$20 times enhanced ferroelectric remanent polarization value compared to that of {\it{x=0}} (also, evident from their room-temperature PE loops, shown in Fig. \ref{PE loop} of the supplementary material below), is also more insulating than {\it{x=0}} (the optical bandgap of {\it{x=0.08}} is greater than that of {\it{x=0}} as determined from diffused reflectance spectroscopy measurements) as shown in Fig.\ref{MultiferroicityImproved}(a). Therefore, {\it{x=0.08}} (Bi-Fe codoped BTO) which is found to have much enhanced ferroelectric, ferromagnetic and insulating properties at room-temperature compared to {\it{x=0}} (only Fe-doped BTO), is more suitable as a better multiferroic. Fig.\ref{MultiferroicityImproved}(b) and its inset show isothermal MH plots of {\it{x=0.08}} and {\it{x=0}} respectively. The difference in 300 K and 10 K MH curves is more prominent in the former than the later. Variations of their extracted coercive fields (H$_C$) and remanent magnetizations (M$_R$) are shown in Fig.\ref{MultiferroicityImproved}(c) and its inset. Here, we note that though M$_R$ increases on decreasing temperature like an usual ferromagnet \cite{XHHuang2008,VMarkovich2010,SPatankar2010}, however, the coercive field shows opposite trend. Even, such surprising behavior is stronger in {\it{x=0.08}} than {\it{x=0}}. It is important to note that such unusual trend of coercive field with temperature has been used as an indicative (indirect) measure of the ME coupling strength in many related ME multiferroic systems \cite{NWang2005,TJPark2010,BAhmmad2016}. The presence of ME coupling in these compounds also becomes evident from the correlation of the ferroelectric polarization and the spontaneous magnetization at various temperatures, as shown in Fig. \ref{ME-coupling} of the supplementary material below, similar to that observed in the ME multiferroic 0.9BiFeO$_3$-0.1BaTiO$_3$ system \cite{ASingh2008}. Thus, such anomalous temperature dependent variations of magnetic coercivity in our Bi-Fe codoped compounds is likely governed by the competition between magnetocrystalline anisotropy and ME coupling \cite{NWang2005,REFayling1978,BRuette2004}. The ratio of H$_C$ at 300 K to H$_C$ at 10 K (the factor of reduction of coercive field), can, thus, be used as an indicative marker for the strength of ME coupling [which is plotted in Fig.\ref{MultiferroicityImproved}(d)], which, for this series, becomes the strongest for {\textit{x=0.05}} composition, possibly driven by the right appropriate mix of simultaneously large ferroelectric polarization and ferromagnetic moment value, likely within the same tetragonal phase.

\section{Summary}

In summary, we have investigated room-temperature multiferroic properties of Ba$_{1-x}$Bi$_{x}$Ti$_{0.90}$Fe$_{0.10}$O$_{3}$ (0$\leq$ {\it{x}} $\leq$0.10) compounds. By codoping Bi and Fe into BaTiO$_3$, we completely recover the ferroelectric tetragonal phase in the magnetically doped BTO compounds. However, the role of sample tetragonality on the recovered ferroelectricity is found to be stronger. Both macro- and microscopic measurements reveal that intrinsic ferromagnetic property gets enhanced up to {\it{x=0.08}} with Bi doping, which is followed up by a reduction for {\it{x=0.10}} due to increased antiferromagnetic interaction. Interestingly, in this series of compounds, though {\it{x=0.02}} shows highest ferroelectric polarization and {\it{x=0.08}} shows largest magnetic moment, however, the signature of ME coupling, as indicated from anomalous change of magnetic coercive field with temperature, seems to be strongest in the intermediate {\it{x=0.05}} due to the simultaneous presence of both ferroelectricity and ferromagnetism in significant amount and likely within the same tetragonal phase. However, it would be interesting to further probe such ME coupling present in these compounds by some direct measurement technique. Thus, simultaneous Bi-Fe codoped BTO compounds around the compositions \textit{x=0.05} to \textit{x=0.08} are found to exhibit better room-temperature ME multiferroic properties compared to only Fe doped BTO as shown in Fig.\ref{Phasediagram}.

\section{Acknowledgement}

We acknowledge the use of XPS under the DST-FIST (India) facility in the Department of Physics, IIT Kharagpur for this work. PP would like to acknowledge the financial support from MHRD, India. DC would like to acknowledge SERB, DST, India (project file no. ECR/2016/000019) and BRNS, DAE (sanction number 37(3)/20/23/2016-BRNS) for financial support. DT would like to acknowledge financial support by DST under the India@DESY collaboration.

\section{Supplementary Material}

\begin{figure*} 
	
	\centering \scalebox{0.6}{\includegraphics{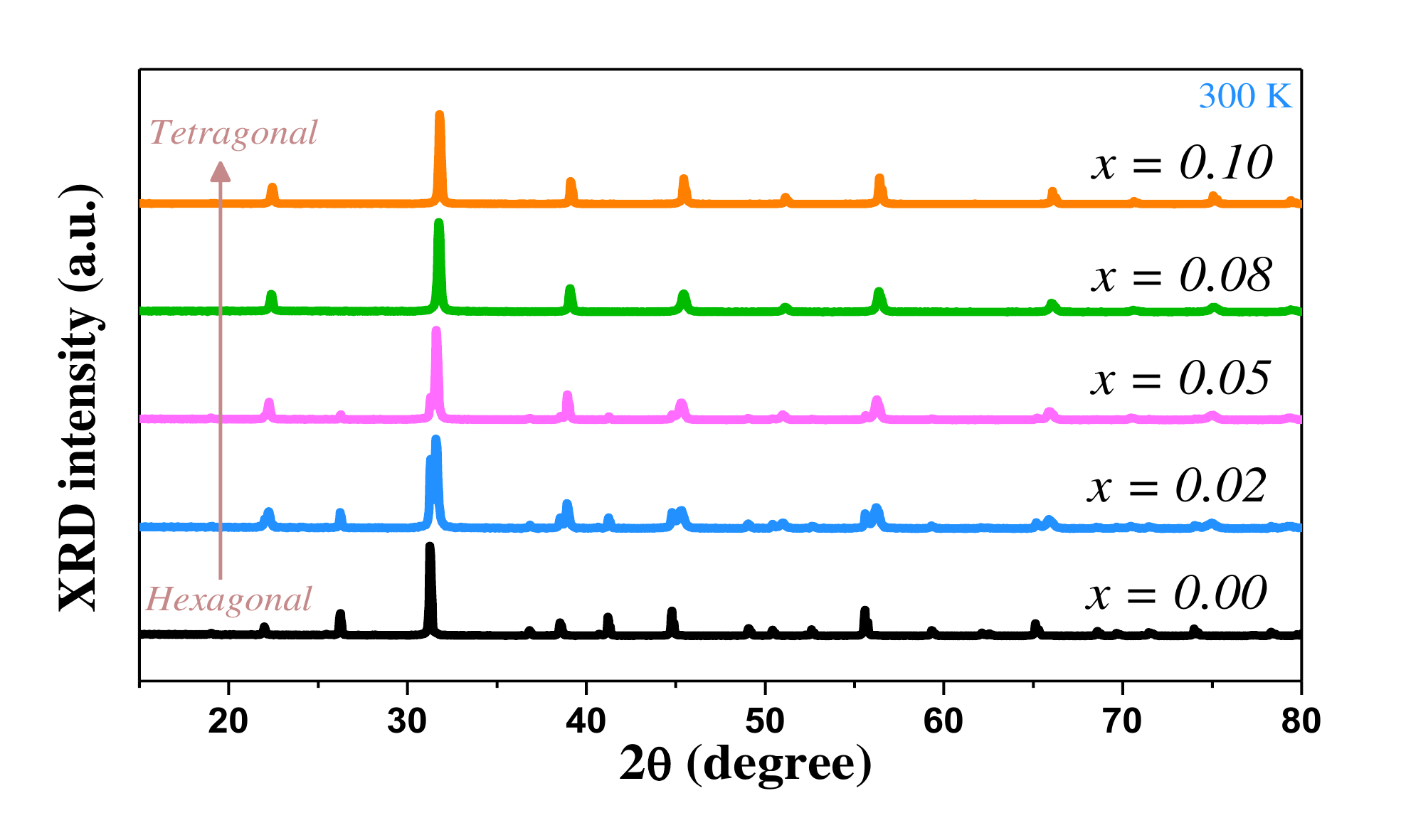}}\caption{Room-temperature XRD spectra of all Ba$_{1-x}$Bi$_{x}$Ti$_{0.90}$Fe$_{0.10}$O$_{3}$ (0$\leq${\it{x}}$\leq$0.10) compounds, which clearly show that with increasing Bi doping concentration paraelectric-hexagonal phase gets suppressed with the concomitant recovery of the ferroelectric tetragonal phase (room temperature XRD spectrum of standard hexagonal BaTiO$_3$ is taken from ICSD database).}\label{BBTFO-XRD}
	
\end{figure*}

\subsection{Room-temperature high-resolution XRD spectra (refer to Fig. \ref{BBTFO-XRD})}

\begin{figure*} 
	
	\centering \scalebox{01}{\includegraphics{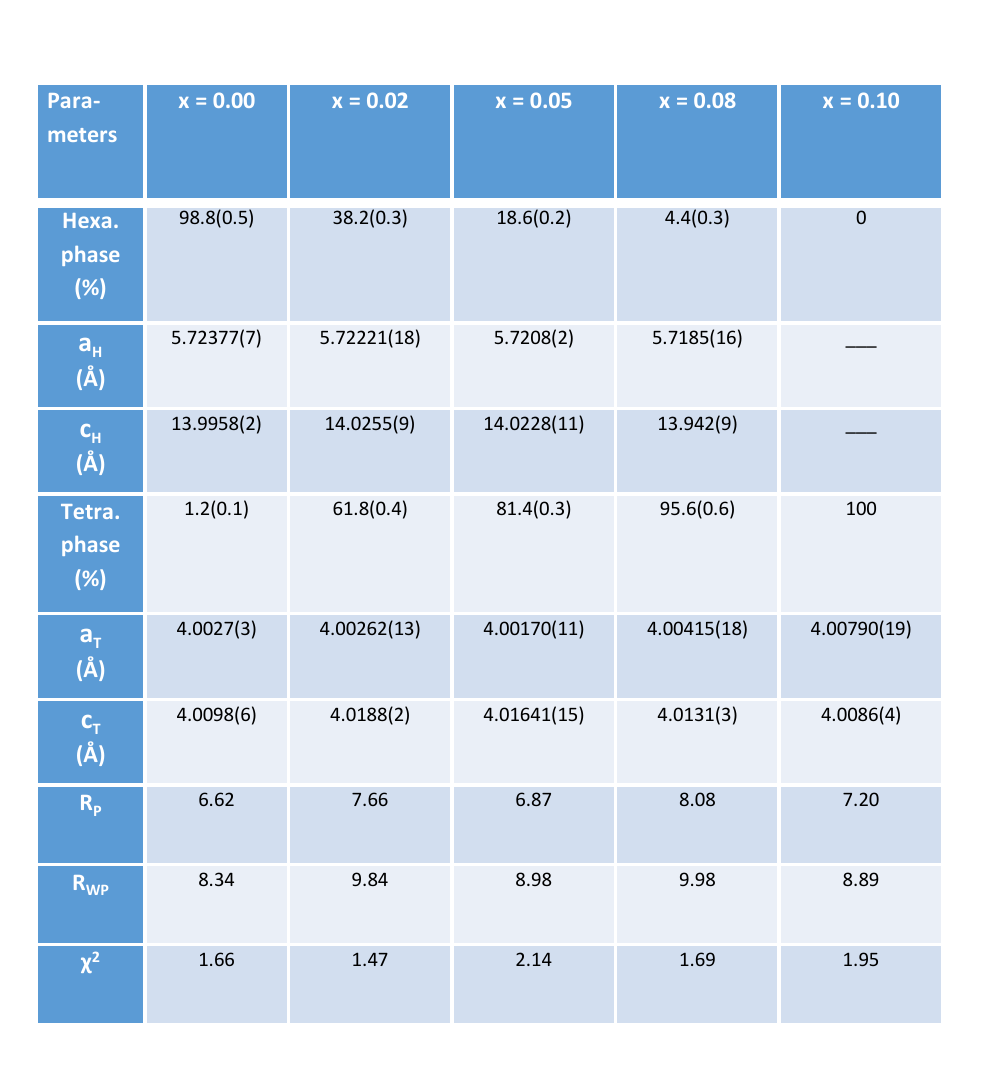}}\caption{Details of structural and Rietveld refinement parameters of all samples.}\label{Refinement-details}
	
\end{figure*}

\subsection{Details of Rietveld refinement (refer to Fig. \ref{Refinement-details})}

\begin{figure*} 
	
	\centering \scalebox{0.8}{\includegraphics{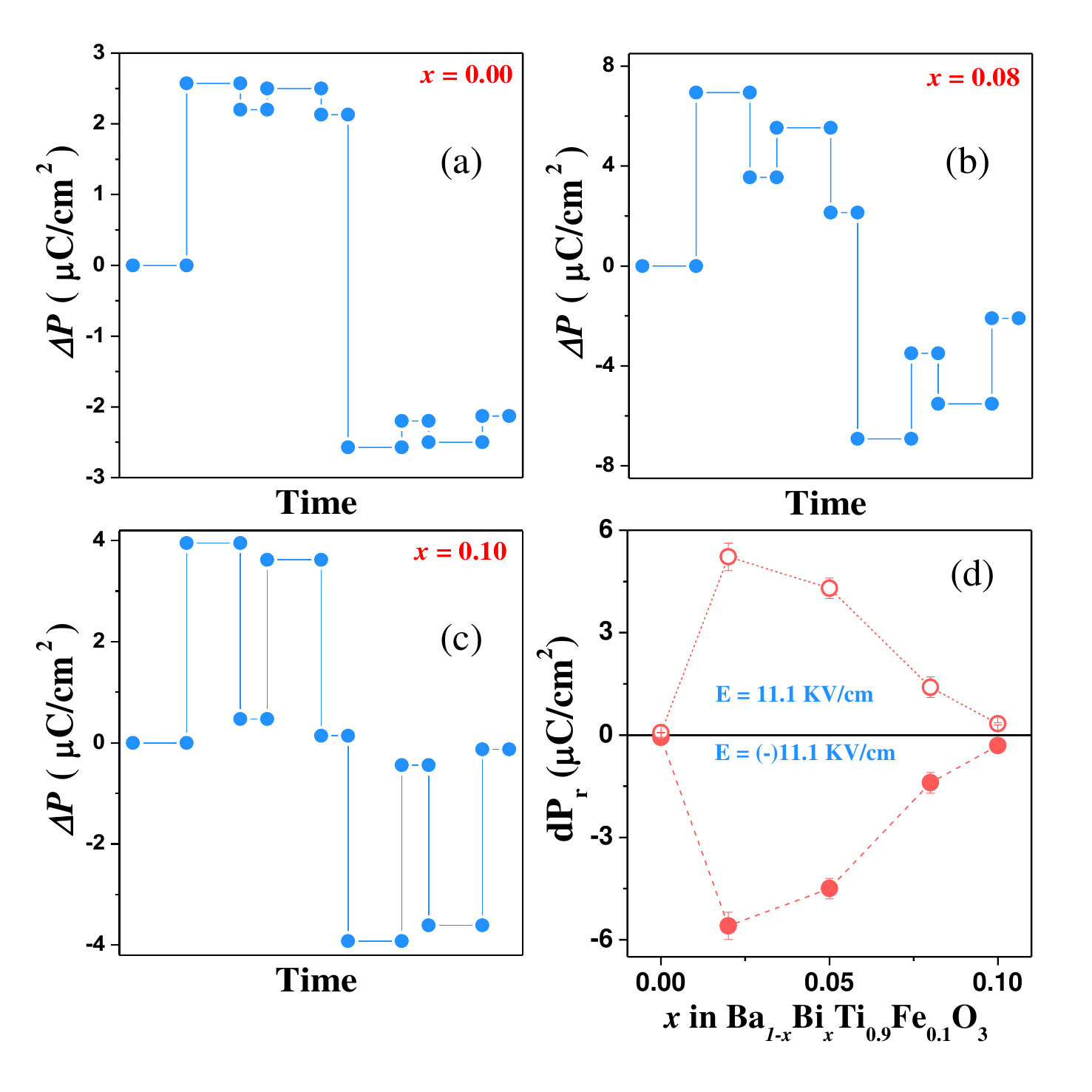}}\caption{(a), (b) and (c) respectively display ferroelectric PUND signals obtained from \textit{x}=0.00 (BaTi$_{0.9}$Fe$_{0.1}$O$_3$), \textit{x}=0.08 (Ba$_{0.92}$Bi$_{0.08}$Ti$_{0.9}$Fe$_{0.1}$O$_{3}$) and \textit{x}=0.10 (Ba$_{0.9}$Bi$_{0.1}$Ti$_{0.9}$Fe$_{0.1}$O$_{3}$) compounds (others are not shown). The applied PUND pulses were of height $\pm$11.1 KV/cm and of pulse width of 100 ms, whereas the delay between two consecutive pulses were 1 ms. For all samples, we find finite presence of intrinsic remanent ferroelectric polarization (dPr) [for example as indicated in (b)]. (d) Shows variation of remanent polarization with increasing Bi concentrations, as obtained from PUND measurements, which gets switched on the reversal of applied electric field, signifies the intrinsic ferroelectric nature of the system.}\label{PUND}
	
\end{figure*}

\subsection{Switching  of  intrinsic ferroelectric remanent polarization as obtained from PUND measurements at room-temperature (refer to Fig. \ref{PUND})}

\begin{figure*} 
	
	\centering \scalebox{0.8}{\includegraphics{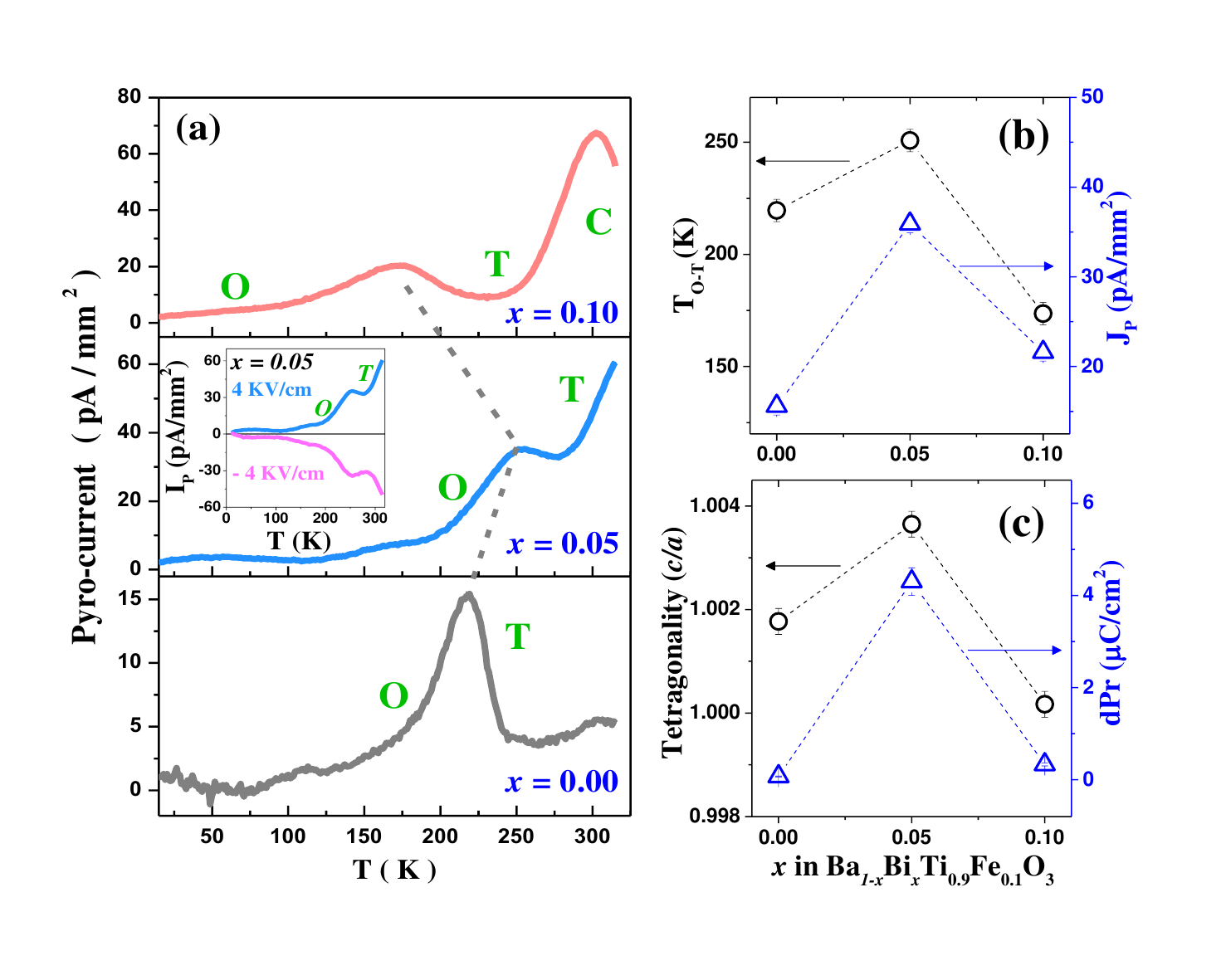}}\caption{(a) Temperature dependent pyroelectric response of  \textit{x}=0.00,  \textit{x}=0.05 and \textit{ x}=0.10 compounds. Inset of (a) shows switching of pyrocurrent for \textit{x}=0.05 (which is also seen for other compounds) on the reversal of poling applied electric field direction. (b) shows variations of  orthorhombic totetragonal phase transition temperature (T$_{O-T}$) and pyrocurrent density (J$_P$) at the transition temperature for these three compounds (\textit{x}=0.00, 0.05 and 0.10). (c) variations of sample tetragonality (\textit{c/a}) and remanent polarization (dPr).}\label{Pyro}
	
\end{figure*}

\subsection{Temperature  dependent  pyroelectric  current  measurements (refer to Fig. \ref{Pyro})}

In pyroelectric measurements, samples were first cooled down to 10 K from 315 K under the presence of applied poling-electric field of 4 KV/cm. Then, the electric field was switched off and the electrodes were short-circuited for sufficient time to get rid of any residual surface charge effects. Sample was then heated to room-temperature at a rate of 10 K/min and temperature dependent pyroelectric current data was recorded. Here, in Fig. \ref{Pyro}(a) we see that \textit{x}=0.00 shows ferroelectric-orthorhombic (\textit{O}) to ferroelectric-tetragonal (\textit{T}) phase transitions at T$_{\textit{O-T}}$ $\sim$218 K during heating due to the presence of minority ferroelectric tetragonal phase.  Now, if we track down this transition for \textit{x}=0.05 and \textit{x}=0.10 (shown by the black dashed line) and note down the change in transition temperature (T$_{\textit{O-T}}$) as well as pyro-current-density (I$_P$) at that point, we see non-monotonic response (they become maximum for \textit{x}=0.05 and then again decrease) as shown in Fig. \ref{Pyro}(b). Such behavior is, however, readily understood when we compare the remanent polarization values (which corresponds to the pyrocurrent density) and room-temperature tetragonality [which decides the magnitude of orthorhombic to tetragonal phase transition temperature (T$_{\textit{O-T}}$)] of these three compounds as shown in Fig. \ref{Pyro}(c). It is also interesting to note that the pyro-response in these compounds gets switched on the reversal of applied poling electric field as shown in the inset of Fig. \ref{Pyro}(a), which further confirms the intrinsic nature of the observed ferroelectricity in these compounds.

\begin{figure*} 
	
	\centering \scalebox{0.8}{\includegraphics{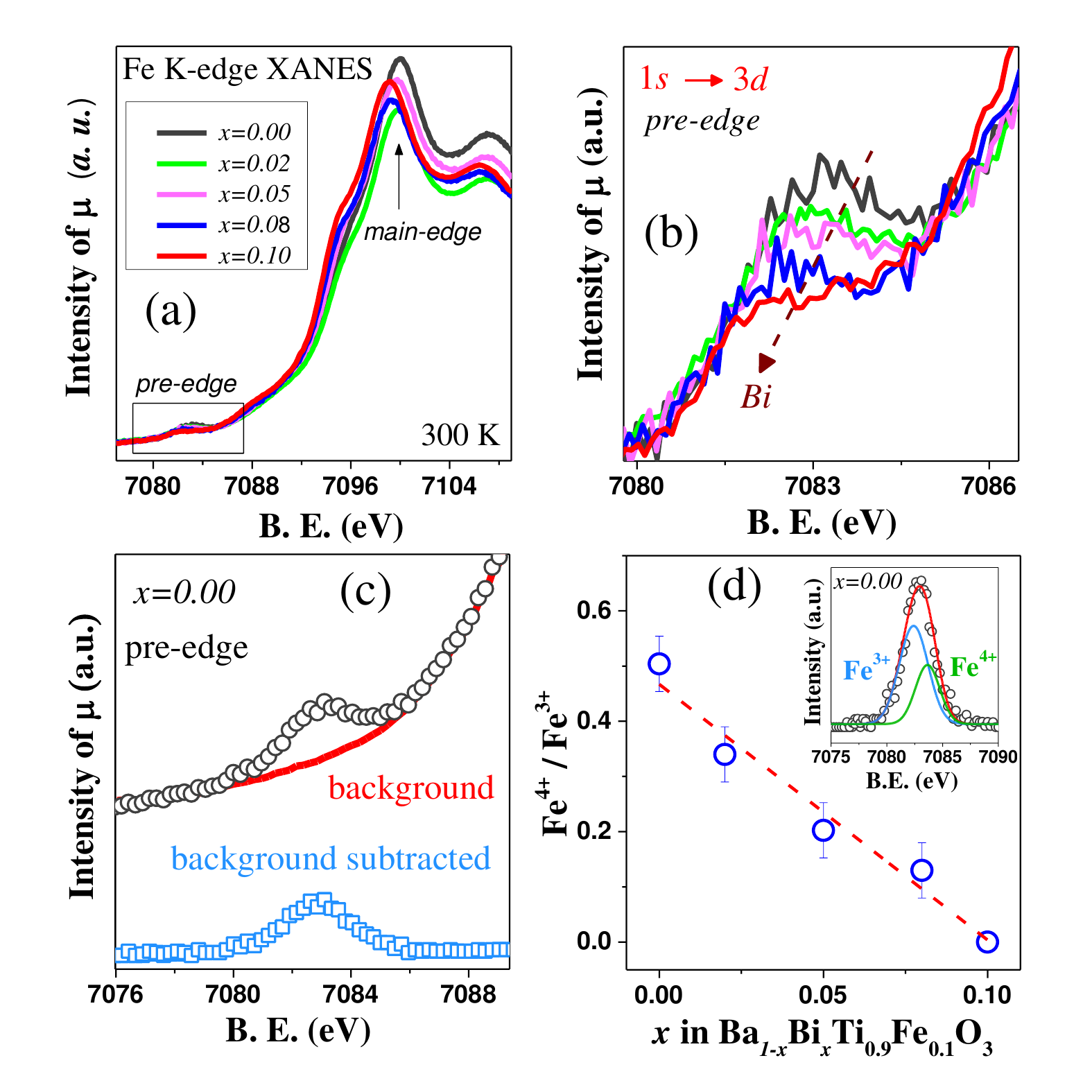}}\caption{(a) Fe K-edge XANES spectra taken at room-temperature. (b) Close-up view of the pre-edge peak for all Bi-Fe codoped BTO compounds. (c) Background subtraction of the pre-edge peak and (d) variation of the Fe4+ to Fe3+ ratio with Bi doping as obtained from two-peaks fitting of the background subtracted pre-edge peak, shown in its inset.}\label{XANES}
	
\end{figure*}

\subsection{Room-temperature Fe \textit{K}-edge XANES spectra of Ba$_{1-x}$Bi$_{x}$Ti$_{0.90}$Fe$_{0.10}$O$_{3}$, 0$\leq${\it{x}}$\leq$0.10 (refer to Fig. \ref{XANES})}

Fe \textit{K}-edge XANES spectra can be subdivided into two regions; pre-edge region [1\textit{s} $\rightarrow$ 3\textit{d} excitation, see Fig. \ref{XANES}(b)] and main edge region (1\textit{s} $\rightarrow$ 4\textit{p} excitation). Both features are observed to shift towards lower energy with increasing Bi doping content, which is an indication that Fe valence state steadily decreases from nominal 4+ to 3+. As all these samples are associated with oxygen-vacancies to some extent, there also remains Fe$^{3+}$. So, to quantify a nominal Fe$^{3+}$ to Fe$^{3+}$ ratio, we investigate the pre-edge peak region. The tentative amounts of nominal Fe$^{3+}$ in these samples are calculated from the background subtraction [see Fig. \ref{XANES}(b)] and subsequent deconvolution of the pre-edge peak into Fe$^{3+}$ and Fe$^{3+}$ contributions as shown in the inset to Fig. \ref{XANES}(d).

\begin{figure*} 
	
	\centering \scalebox{0.8}{\includegraphics{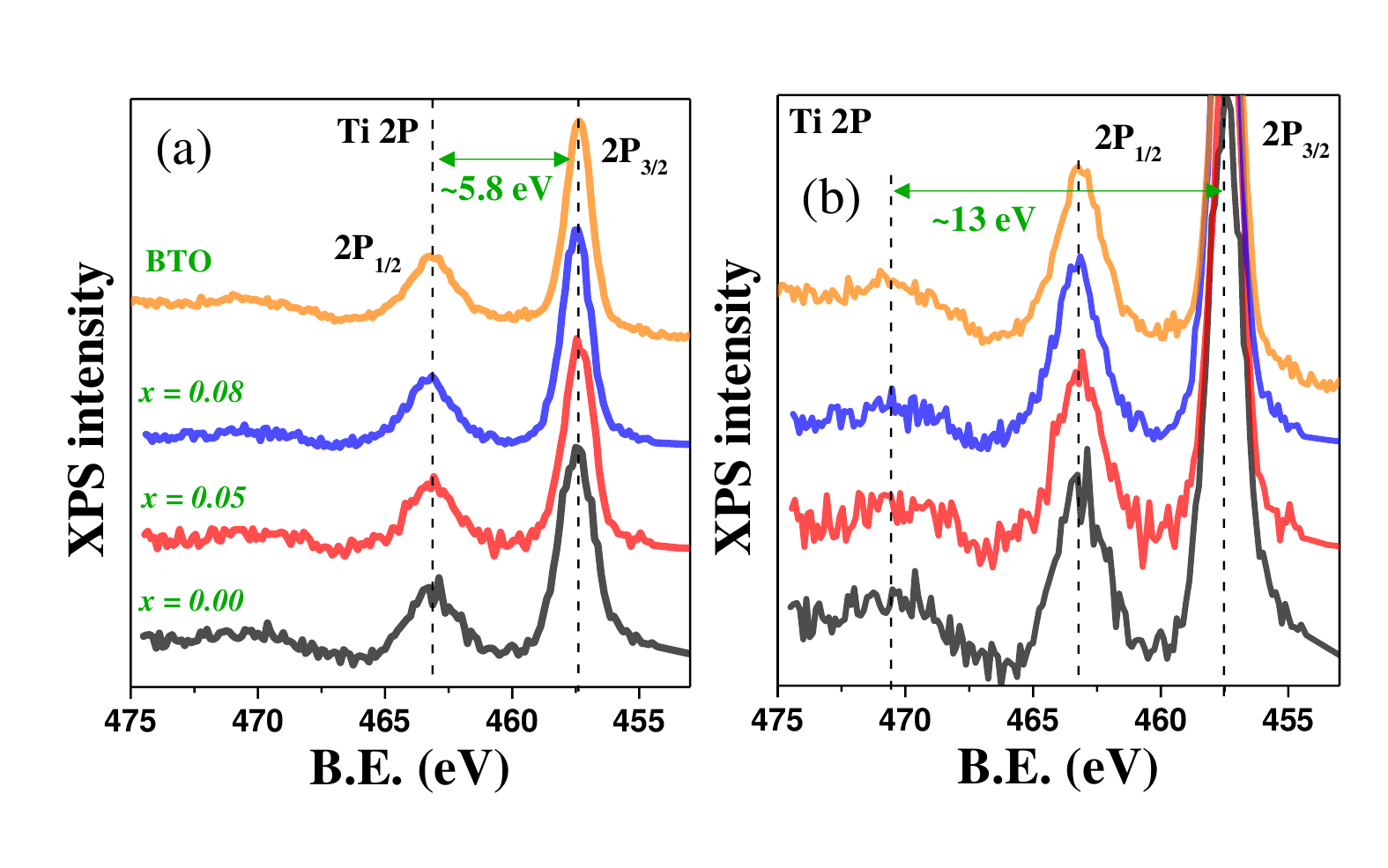}}\caption{(a) Ti 2\textit{p} corelevel XPS spectra taken at room-temperature shows the energy difference between 2P$_{3/2}$ and 2P$_{1/2}$ is around $\sim$5.8 eV which is a characteristic of the presence of Ti$^{4+}$ valence state (b) Close-up view of the satellite features of Ti 2p XPS spectra, which are at $\sim$13 eV away from the main 2P$_{3/2}$  peak. This is also a signature of Ti$^{4+}$ valence state.}\label{XPS}
	
\end{figure*}

\subsection{Room-temperature Ti-2\textit{p} core level XPS spectra of Ba$_{1-x}$Bi$_{x}$Ti$_{0.90}$Fe$_{0.10}$O$_{3}$, 0$\leq${\it{x}}$\leq$0.10  (refer to Fig. \ref{XPS})}

\subsection{Tolerance  factor  calculations of Ba$_{1-x}$Bi$_{x}$Ti$_{0.90}$Fe$_{0.10}$O$_{3}$, 0$\leq${\it{x}}$\leq$0.10}

Goldschmidt’s tolerance factor which provides an effective and simple way of investigating structural phase stabilities of ABO$_3$ kind of perovskite, can be quantified through the knowledge of ionic radii for a particular coordination number and valence information. Here, we have determined the tolerance factor of all Bi and Fe codoped BTO compounds using Fe valence states derived from XANES and Ti valence state from XPS studies. We have also noted that Ba and Bi remains in their usual valences as observed in XPS (not shown here). The formula for Goldschmidt’s tolerance (GT) factor is given by
\begin{equation}
GT=\frac{r_A+r_O}{\sqrt{2}(r_B+r+O)}          
\end{equation}

Where r$_A$, and r$_B$ are ionic radii of the A and B site cations , while r$_O$ is that of oxygen anion. Now, as A and B site cations (Ba and Ti) are simultaneously doped by other cations (Bi and Fe), we may rewrite the above equation for the determination of tolerance factor of the compound Ba$_{1-x}$Bi$_{x}$Ti$_{0.90}$Fe$_{0.1-y}$$^{3+}$Fe$_{y}$$^{4+}$O$_{3}$ (Fe remains in both 3+ and 4+ states).
\begin{equation}
GT=\frac{(1-x)r_{Ba} + x r_{Bi} + r_O}{\sqrt{2}[(0.1-y)r_{Fe^{3+}} + y r_{Fe^{4+}} + r_O)}          
\end{equation}

Where \textit{x} is the Bi doping concentration and \textit{y} is the Fe$^{4+}$ content which is determined from XANES study [see Fig. \ref{XANES}(d)]. We have used the Shanon table for the information of ionic radii of the following ions; Ba$^{2+}$(XII) = 1.61 \AA, Bi$^{3+}$(XII) = 1.45 \AA, Ti$^{4+}$(VI) = 0.605 \AA,  Fe$^{3+}$(VI) = 0.645 \AA, Fe$^{4+}$(VI) = 0.585 \AA and O$^{2-}$(VI) = 1.4 \AA, where the bracketed number denotes the corresponding coordination number.

\begin{figure*} 
	
	\centering \scalebox{0.8}{\includegraphics{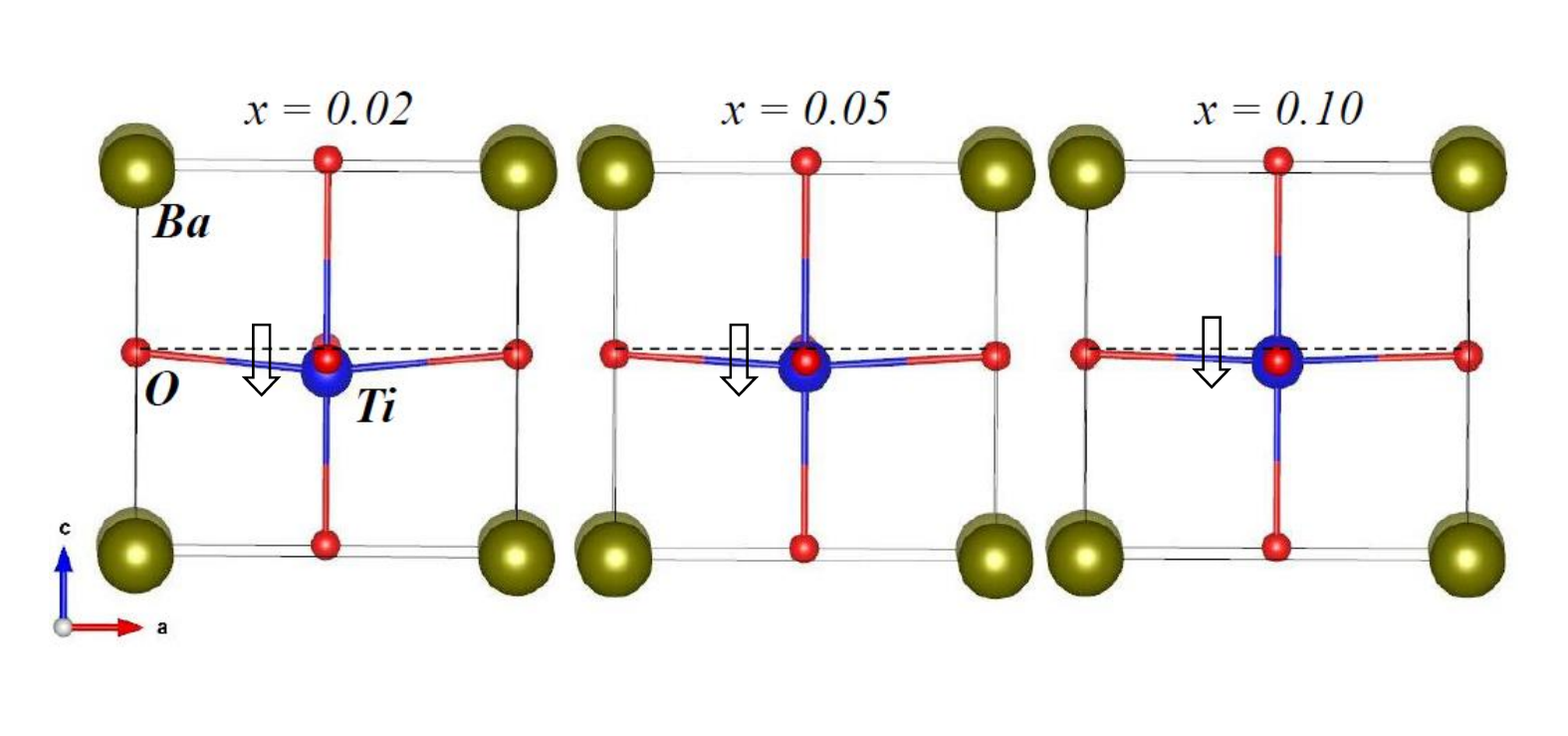}}\caption{Schematic representation of the Ti off-centric distortions (shown by the arrow) in the octahedral environment of the tetragonal BaTiO3 phase,which is based on data from our Rietveld refinement results. Such off-centric distortion directly reflects the observed sample tetragonality (\textit{c/a}) and ferroelectric polarization. Here, we note that the distortion gradually decreases from \textit{x}=0.02 as Bi doping concentration is increased.}\label{Octahedral-distortion}
	
\end{figure*}

\subsection{Schematic  visualization of  octahedral distortion  of Ba$_{1-x}$Bi$_{x}$Ti$_{0.90}$Fe$_{0.10}$O$_{3}$, 0$\leq${\it{x}}$\leq$0.10 (refer to Fig. \ref{Octahedral-distortion})}

\begin{figure*} 
	
	\centering \scalebox{0.8}{\includegraphics{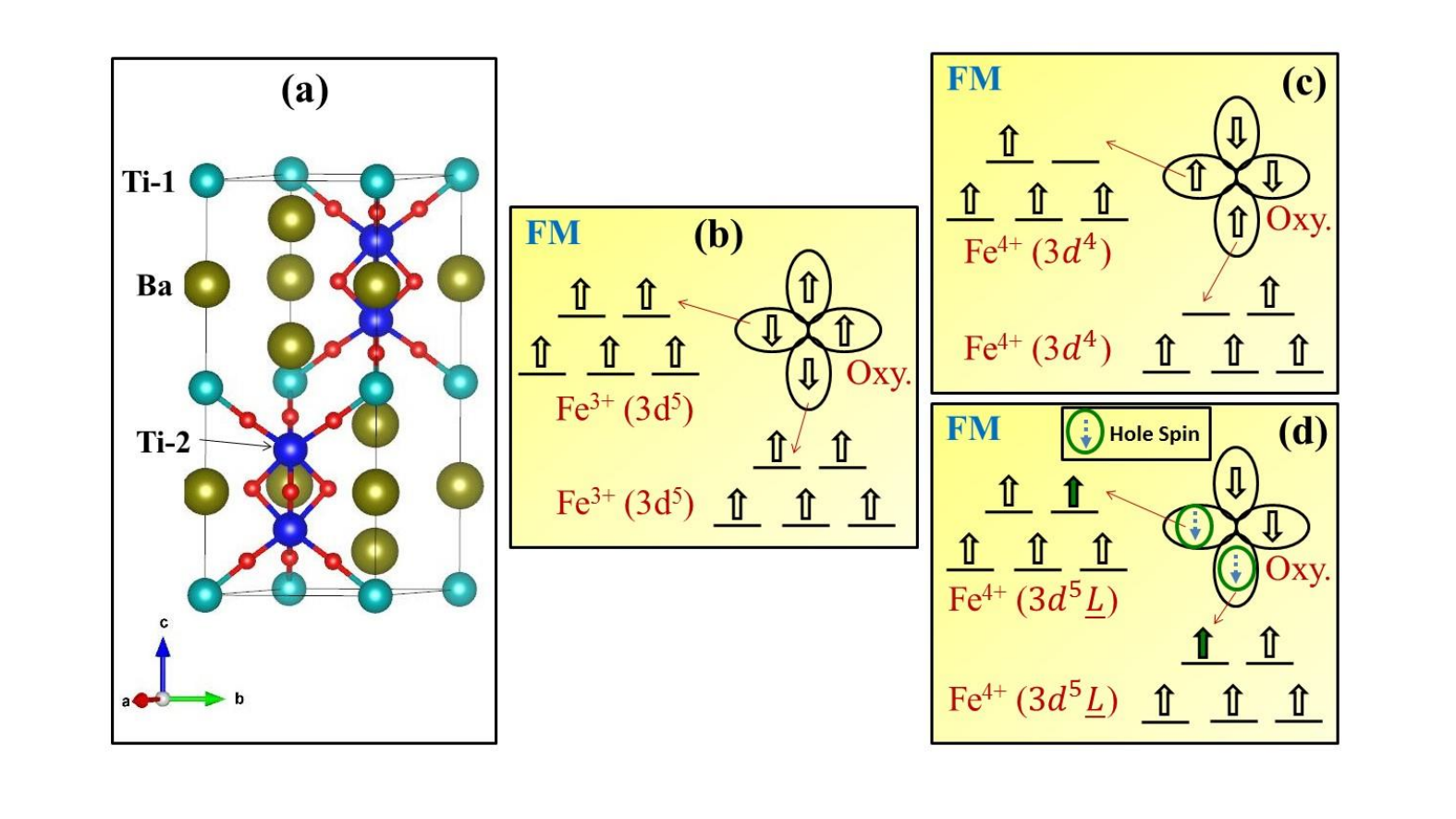}}\caption{(a) Unit cell crystal structure of BaTiO3 hexagonal polymorph. (b) and (c) represent possible 90$^0$ ferromagnetic superexchange interactions in the hexagonal phase between two adjacent Fe$^{3+}$ ions and Fe$^{4+}$ (considering the ionic 3d$^4$ configuration) ions respectively. Fe$^{4+}$, because of its high valence-state, is also associated with some 3d$^5$$\underbar{L}$ contribution to its ground-state wave function ($\underbar{L}$ denotes an oxygen hole left by the electron transfer into Fe) and (d) indicates the ferromagnetic alignment of the holes left on the two oxygen lobes, arising from such charge transfer phenomenon.}\label{Hexagonal-superexchange}
	
\end{figure*}

\subsection{Schematic representation of the magnetic superexchange interactions in Bi-Fe dual doped BTO compounds (refer to Fig. \ref{Hexagonal-superexchange} and Fig. \ref{Tetragonal-superexchange})}

Hexagonal BTO has two different Ti sites as shown in the following Fig. \ref{Hexagonal-superexchange}(a). Fe ions can substitute both type of Ti ions. Now, the interaction between Fe-1 (Fe replacing Ti-1) and Fe-2 (Fe replacing Ti-2) as well as between Fe-1 ions are very weak and give rise to large paramagnetic background. But interestingly, the interaction between Fe-2 ions are strong ferromagnetic with a Curie temperature of T$_C$$>$600 K. The superexchange interactions possible to give rise ferromagnetism in the hexagonal phase are shown in Fig. \ref{Hexagonal-superexchange}(b) and (c). Again, two Fe-2 ions show strong tendency to cluster and can have interaction through direct overlap of atomic orbitals.

Whereas on the contrary, in the tetragonal BTO phase, the possible exchange that can give rise to ferromagnetism is 180$^0$ superexchange between Fe$^{3+}$ and Fe$^{4+}$ ions as shown in Fig. \ref{Tetragonal-superexchange}(a). As there is only one type of Ti sites in the tetragonal BTO, Fe ions without nearest neighbor other Fe ions can give rise to paramagnetism. Hence, the amount of paramagnetic background in the higher Bi doped samples where majority is tetragonal phase, is smaller than that of the lower or without Bi doped samples, where hexagonality is more. Also, as 180$^0$ ferromagnetic superexchange interaction is much stronger than the corresponding 90$^0$ ferromagnetic exchange, higher Bi doped sample exhibits larger magnetic moment as well as higher ferromagnetic TC. However, for \textit{x}=0.10 where all Fe seems to be in 3+ (3d$^5$) state, major 1800 super exchange interaction is antiferromagnetic as shown in Fig. \ref{Tetragonal-superexchange}(b). Hence, beyond \textit{x}=0.08, there is a drop in the magnetic moment for \textit{x}=0.10 compound.

\begin{figure*} 
	
	\centering \scalebox{0.8}{\includegraphics{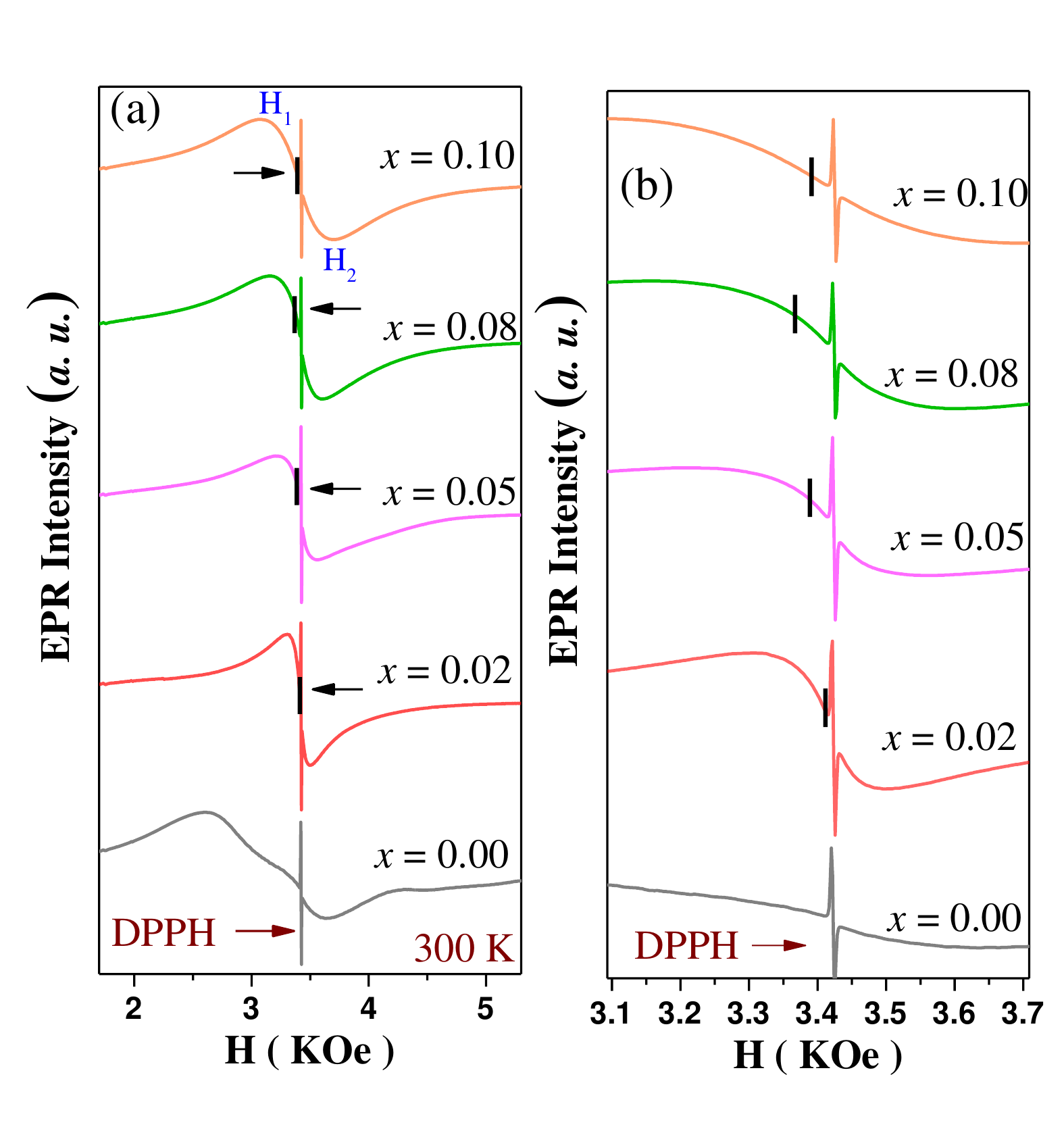}}\caption{(a) Room-temperature X-band (~9.60 GHz) EPR spectra of Ba$_{1-x}$Bi$_{x}$Ti$_{0.90}$Fe$_{0.10}$O$_{3}$ (0$\leq${\it{x}}$\leq$0.10). The black line indicated by the arrow shows the position of resonance magnetic field. (b) Shows zoomed close-up view of the variation of resonance position with increasing Bi doping concentration.}\label{EPR}
	
\end{figure*}

\subsection{Room-temperature Electron Paramagnetic Resonance (EPR) spectra of Ba$_{1-x}$Bi$_{x}$Ti$_{0.90}$Fe$_{0.10}$O$_{3}$, 0$\leq${\it{x}}$\leq$0.10 (refer to Fig. \ref{EPR})}

The following room-temperature EPR spectra shows that while \textit{x}=0.00 is composed of a broad asymmetric line, possibly consisting of two or more spectra, the rest of the doped compounds rather exhibit more clean and symmetric spectra. Here, the position of resonance magnetic field is calculated by H$_{Res.}$=(H$_1$+H$_2$)/2, where H$_1$ and H$_2$ are magnetic field values of the maximum and minimum intensity points of the corresponding resonance spectrum. We see that from \textit{x}=0.02 as Bi doping concentration is increased resonance field value decreases up to \textit{x}=0.08, and then again increases slightly for \textit{x}=0.10 as indicated in the following Fig. \ref{EPR}.

\begin{figure*} 
	
	\centering \scalebox{0.8}{\includegraphics{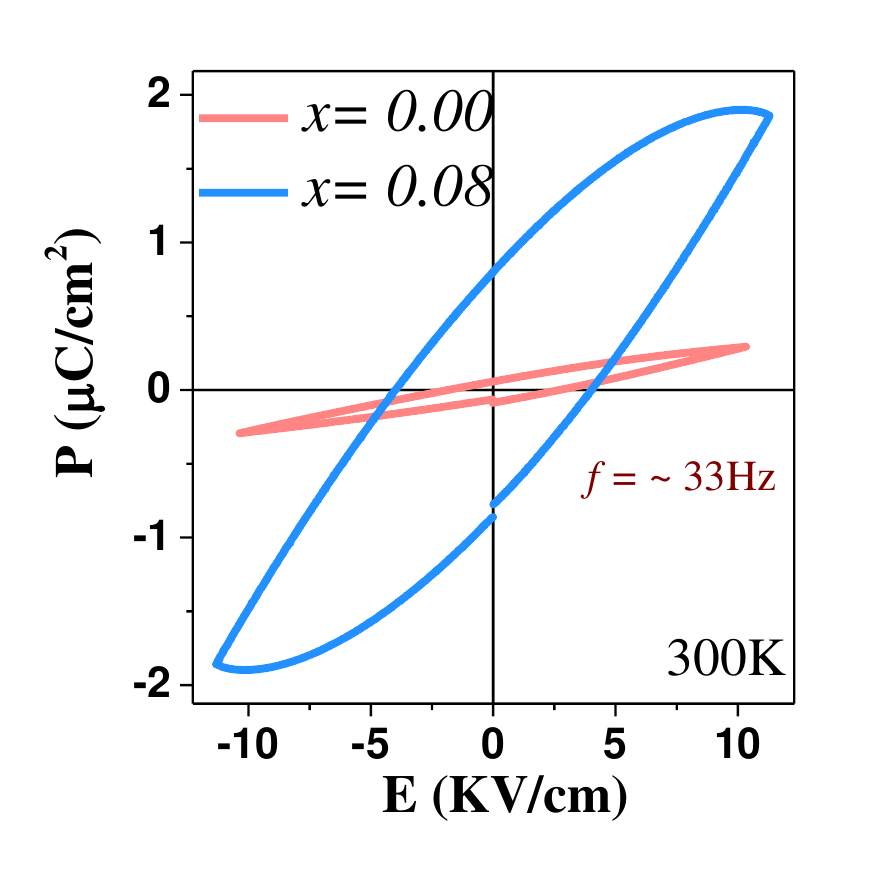}}\caption{Room-temperature PE loops of \textit{x}=0.00 (BaTi$_{0.9}$Fe$_{0.1}$O$_{3}$) and \textit{x}=0.08 (Ba$_{0.92}$Bi$_{0.08}$Ti$_{0.9}$Fe$_{0.1}$O$_{3}$) taken at a frequency of $\sim$33 Hz. Here, \textit{x}=0.00 shows cigar shaped PE loop indicating lossy nature of the sample, while \textit{x}=0.08 shows improved and robust ferroelectricity (which is also observed in PUND measurements as shown in Fig. \ref{PUND}).}\label{PE loop}
	
\end{figure*}

\subsection{Room temperature P vs. E loops (refer to Fig. \ref{PE loop})}

\begin{figure*} 
	
	\centering \scalebox{0.8}{\includegraphics{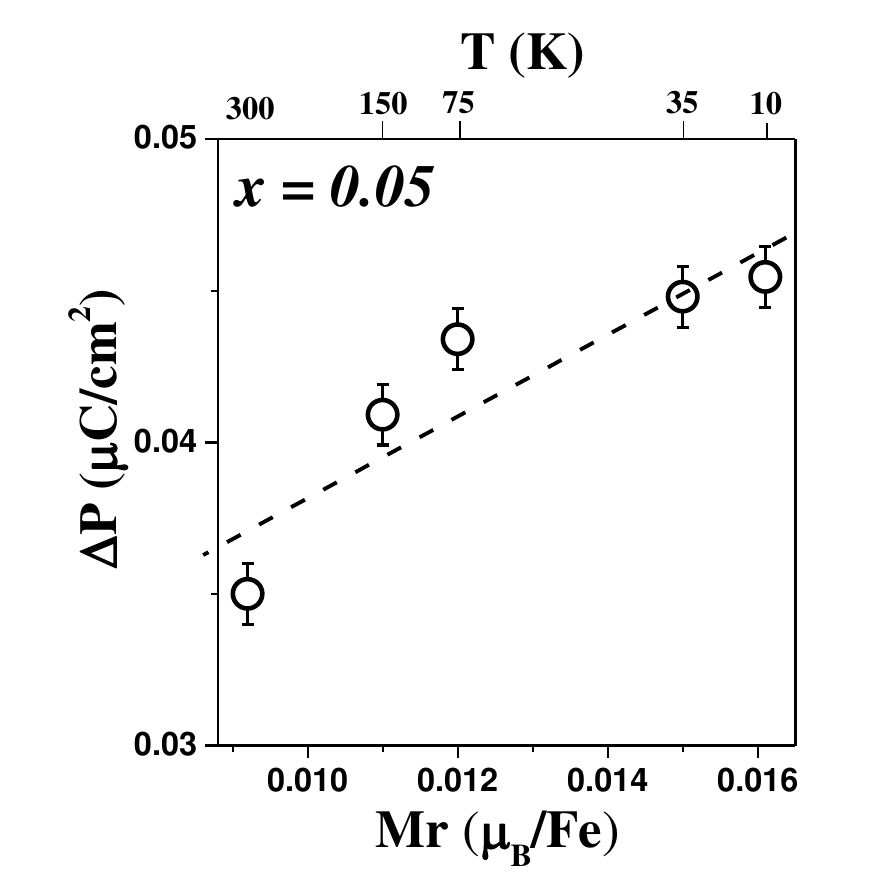}}\caption{Variations of the ferroelectric polarization (obtained by integrating the pyrocurrent density data of \textit{x}=0.05, shown in Fig. \ref{Pyro}) with the remanent magnetization (bottom \textit{x}-axis) at the corresponding temperature (top \textit{x}-axis).}\label{ME-coupling}
	
\end{figure*}

\subsection{Correlation of ferroelectric polarization and ferromagnetic remanent moment (refer to Fig. \ref{ME-coupling})}

The evidence of magnetoelectric coupling in these Bi-Fe codoped BTO samples is clear from the following plot of polarization vs. remanent magnetization, where we see that there is a correlation between these two ferroic-orders by same way as previously shown in A. Singh et al. \textit{Phys. Rev. Lett.} \textbf{101}, 247602 (2008), on a similar kind of 0.9BiFeO$_3$-0.1BaTiO$_3$ system. Here, the temperature dependent polarization is deduced by integrating the pyrocurrent vs. temperature data (shown in Fig. \ref{Pyro}) [$\Delta$P=(dt/dT)$\int$(J$_P$ dT), in our case dt/dT was 0.1 sec/K] and remanent magnetization was obtained from isothermal MH measurements at those corresponding temperatures. Thus, these Bi-Fe doped BTO compounds are promising for room-temperature magnetoelectric-multiferroicity.

\subsection{Role of extrinsic effects on the observed  magnetism}

Contributions from extrinsic impurity phases to the observed magnetic response have often been observed in dilute magnetic semiconductor systems. Here, we have, thus, tried to analyze any possible magnetic contributions from possible impurity phases (like BiFeO$_3$, Fe$_2$O$_3$, FeTiO$_3$ and Fe$_3$O$_4$) in our Bi-Fe co-doped BTO samples. Notably, from room-temperature high-resolution x-ray diffraction (XRD) and micro-Raman spectroscopy we could not detect any impurity phases in our Bi-Fe dual doped BTO compounds. Further, as detailed below, the trend observed in our magnetic data cannot be understood from the presence of trace quantities of any such possible impurity phases, which could have gone undetected in our detailed structural investigations.

\textbf{BiFeO$_3$}: It has a paramagnetic to antiferromagnetic phase transition temperature at T$_N$ $\sim$645 K. From our results, we see that \textit{x}=0.00 (BaTi$_{0.9}$Fe$_{0.1}$O$_{3}$) has a transition temperature T$_C$ $\sim$ 675 K, but as there is no Bi ion in these compound, there cannot be any role of BiFeO$_3$ to the observed magnetic response. Further, for higher Bi doped sample such as \textit{x}=0.08 (Ba$_{0.92}$Bi$_{0.08}$Ti$_{0.9}$Fe$_{0.1}$O$_3$) T$_C$ is $\sim$ 770 K, much higher than that of BiFeO$_3$. So, the contribution of BiFeO$_3$ in the Bi-Fe dual doped BTO compounds is further ruled out.

\textbf{Fe$_2$O$_3$ or Fe$_3$O$_4$}:Fe$_2$O$_3$ (hematite) has a Neel temperature $\sim$ 950 K, which is much higher than the observed magnetic transition in our samples, so its contribution to the observed magnetic response can be ruled out. Fe$_3$O$_4$ has a paramagnetic to ferromagnetic phase transition around $\sim$850 K, which again is much higher than the highest transition temperature ($\sim$770 K) observed for \textit{x}=0.08 (Ba$_{0.92}$Bi$_{0.08}$Ti$_{0.9}$Fe$_{0.1}$O$_3$) in our Bi-Fe dual doped BTO system. Thus, any contribution of Fe$_3$O$_4$ impurity phase to the observed magnetic response can also be ruled out. 

\textbf{Fe$_2$O$_3$-FeTiO$_3$ solid solution (Ti doped Fe$_2$O$_3$)}: While both Fe$_2$O$_3$ and FeTiO$_3$ are antiferromagnetic (T$_N$ $\sim$65 K for FeTiO$_3$), in their solid solution \textit{y}Fe$_2$O$_3$–(1-\textit{y})FeTiO$_3$, as \textit{y} (Fe$_2$O$_3$) content is varied, the magnetization value goes through a maximum value in the composition range between \textit{y}=0.20 to \textit{y}=0.30. The corresponding ferrimagnetic transition temperatures for \textit{y}=0.20 and 0.30 compositions is in the 250 K - 350 K range. In the Bi-Fe dual-doped BaTiO$_3$ series, while the magnetization value peaks around \textit{x}=0.08 composition, the corresponding magnetic transition temperature is much higher ($\sim$770 K), which cannot be accounted using an impurity magnetic phase arising from a solid-solution between Fe$_2$O$_3$ and FeTiO$_3$.

\textbf{Fe$_3$O$_4$ –FeTi$_2$O$_4$  solid solution}: In Fe$_3$O$_4$ –FeTi$_2$O$_4$ (FeFe$_{2-x}$Ti$_x$O$_4$) solid solution, with decreasing Ti concentration, both Curie temperature as well as saturation magnetization monotonically increases. This is clearly inconsistent with the non-monotonic dependence of magnetization value observed with progressive Bi doping in our samples.

%------------Fig. S1

%------------Fig. S2

%------------Fig. S3

%------------Fig. S4

%------------Fig. S5

%------------Fig. S6

%------------Fig. S7

%------------Fig. S8

%------------Fig. S9

\begin{figure*} 
	
	\centering \scalebox{0.8}{\includegraphics{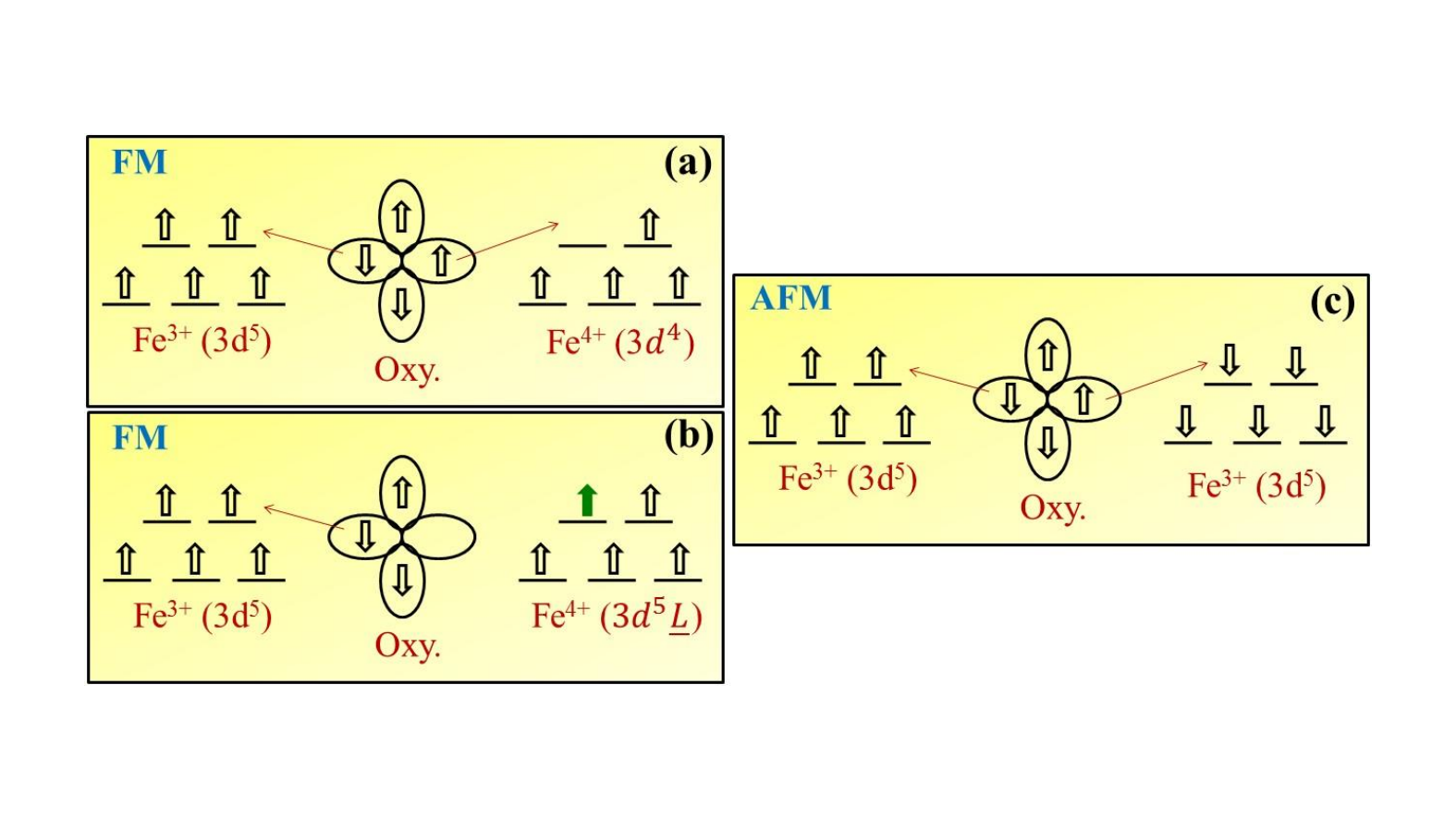}}\caption{(a) and (c) represent possible 180$^0$ ferromagnetic superexchange [between adjacent Fe$^{3+}$ and Fe$^{4+}$ (considering ionic 3d$^4$ configuration) ions for \textit{x}=0.08 ] and antiferromagnetic superexchange [between two adjacent Fe$^{3+}$ ions for \textit{x}=0.10] interactions respectively in the tetragonal phase. (b) represents the ferromagnetic alignment between neighboring Fe$^{3+}$ and Fe$^{4+}$ (considering the 3d$^5$$\underbar{L}$ contribution to its ground state wave function [further details in caption of Fig. \ref{Hexagonal-superexchange}]) ions in the tetragonal structure for \textit{x}=0.08.}\label{Tetragonal-superexchange}
	
\end{figure*}

%------------Fig. S10

%------------Fig. S11

%------------Fig. S12

\end{document}